\documentclass[twocolumn]{aastex62}

\received{2018 November 7th}
\revised{2019 January 31st}
\accepted{2019 February 11th}
\usepackage{graphicx}
\usepackage{amssymb}
\usepackage{amsmath}
\usepackage{epstopdf}
\shorttitle{Hot and cold populations of planetesimals in the $\beta$ Pictoris belt}
\shortauthors{Matr\`a et al.}

\begin{document}


\title{Kuiper Belt-Like Hot and Cold Populations of Planetesimal Inclinations in the $\beta$ Pictoris Belt Revealed by ALMA}

\author[0000-0003-4705-3188]{L. Matr\`a}
%\altaffiltext{1}{Institute of Astronomy, University of Cambridge, Madingley Road, Cambridge CB3 0HA, UK}
\altaffiliation{Submillimeter Array (SMA) Fellow}
\affil{Harvard-Smithsonian Center for Astrophysics, 60 Garden Street, Cambridge, MA 02138, USA}
\email{luca.matra@cfa.harvard.edu}
\author{M. C. Wyatt}
\affil{Institute of Astronomy, University of Cambridge, Madingley Road, Cambridge CB3 0HA, UK}
\author{D. J. Wilner}
\affil{Harvard-Smithsonian Center for Astrophysics, 60 Garden Street, Cambridge, MA 02138, USA}
\author{W. R. F. Dent}
\affil{ALMA Santiago Central Offices, Alonso de Cordova 3107, Vitacura, Santiago, Chile}
\author{S. Marino}
\affil{Max Planck Institute for Astronomy, K\"onigstuhl 17, 69117 Heidelberg, Germany}
%\author{friends}
\author{G. M. Kennedy}
\affil{Department of Physics, University of Warwick, Gibbet Hill Road, Coventry, CV4 7AL, UK}
\author{J. Milli}
\affil{European Southern Observatory (ESO), Alonso de C\'ordova 3107, Vitacura, Casilla 19001, Santiago, Chile}

\begin{abstract}
The inclination distribution of the Kuiper belt provides unique constraints on its origin and dynamical evolution, motivating vertically resolved observations of extrasolar planetesimal belts. We present ALMA observations of millimeter emission in the near edge-on planetesimal belt around $\beta$ Pictoris, finding that the vertical distribution is significantly better described by the sum of two Gaussians compared to a single Gaussian. This indicates that, as for the Kuiper belt, the inclination distribution of $\beta$ Pic's belt is better described by the sum of dynamically hot and cold populations rather than a single component. The hot and cold populations have RMS inclinations of 8.9$^{+0.7}_{-0.5}$ and 1.1$^{+0.5}_{-0.5}$ degrees. We also report that an axisymmetric belt model provides a good fit to new and archival ALMA visibilities, and confirm that the midplane is misaligned with respect to $\beta$ Pic b's orbital plane. However, we find no significant evidence for either the inner disk tilt observed in scattered light and CO emission or the South-West/North-East (SW/NE) asymmetry previously reported for millimeter emission. Finally, we consider the origin of the belt's inclination distribution. Secular perturbations from $\beta$ Pic b are unlikely to provide sufficient dynamical heating to explain the hot population throughout the belt's radial extent, and viscous stirring from large bodies within the belt alone cannot reproduce the two populations observed. This argues for an alternative or additional scenario, such as planetesimals being born with high inclinations, or the presence of a `$\beta$ Pic c' planet, potentially migrating outwards near the belt's inner edge.% through resonance sweeping and scattering.}% or resonance sweeping and scattering by an outward-migrating `$\beta$ Pic c' planet near the inner edge of the belt.

\end{abstract}


%% subject header.  Objects should be in the appropriate "individual"
%% headers (e.g. quasars: individual, stars: individual, etc.) with the
%% additional provision that the total number of headers, including each
%% individual object, not exceed six.  The \objectname{} macro, and its
%% alias \object{}, is used to mark each object.  The macro takes the object
%% name as its primary argument.  This name will appear in the paper
%% and serve as the link's anchor in the electronic edition if the name
%% is recognized by the data centers.  The macro also takes an optional
%% argument in parentheses in cases where the data center identification
%% differs from what is to be printed in the paper.

%\keywords{globular clusters: general ---
%globular clusters: individual(\objectname{NGC 6397},
%\object{NGC 6624}, \objectname[M 15]{NGC 7078},
%\object[Cl 1938-341]{Terzan 8})}



\section{Introduction}
\label{sect:intro}

The spatial distribution of dust in planetesimal belts around main-sequence stars provides significant constraints on the formation and dynamical evolution of planetary systems, both individually and as a population \citep[e.g.][]{Marino2018b, Matra2018b}.
Due to its vicinity and youth, the A6-type main sequence star $\beta$ Pictoris has long been considered a unique laboratory to understand the outcome of the planet formation process. Over more than three decades since the discovery of an infrared excess above the stellar emission \citep{Aumann1985}, indicative of the presence of a dusty disk, numerous studies have been attempting to unravel the complexity of this planetary system. %Though this has led to exciting discoveries, many puzzles remain yet to be solved. 

At only 19.44 pc \citep[][]{vanLeeuwen2007}, the disk's brightness and on-sky extent have made this system extremely suitable for resolved imaging. It was as early as 1984 when the belt's near edge-on geometry was discovered through optical coronagraphic observations \citep{Smith1984}, which showed starlight scattered off micron sized particles extending out to beyond 1000 AU. Soon after, improved scattered light imaging of this system from the ground at optical and near-infrared (near-IR) wavelengths unveiled a complex disk structure with several asymmetries \citep[][]{KalasJewitt1995}. %1) A brightness/ radial size asymmetry between the NE and SW sides of the disk, caused by a steeper surface brightness decrease with projected radius on the SW compared to the NE side. 2) A width asymmetry beyond 7$\arcsec$, with the disk flaring more on the SW than on the NE sides. 3) a wing-tilt asymmetry, with the position angle (PA) of the NE side differing from that on the SW side. 4) a butterfly asymmetry, due to the angles between isophotes and disk midplane being larger in the S than in the N on the NE side of the disk, and vice versa on on the SW side of the disk. 

%A logical explanation of the wing-tilt asymmetry was found already by \citet{KalasJewitt1995} to be a non-isotropic scattering phase function of the grains, where a forward-scattering peak causes material above the disk sky-projected midplane to appear brighter because closer to Earth. This was taken as proof that the disk is not perfectly edge-on, being inclined by an angle between 2$^{\circ}$ and 5$^{\circ}$ to the line-of-sight. This wing-tilt asymmetry was confirmed in all follow-up scattered light observations, and attributed to slight inclinations of $\sim$2$^{\circ}$-6$^{\circ}$ \citep{Ahmic2009}, $\sim$4$^{\circ}$ \citep{Milli2014}, and $\sim$5$^{\circ}$ \citep{Millar-Blanchear2015}, depending on disk structure and phase function assumptions taken in the modelling.

The near edge-on configuration makes $\beta$ Pic's belt extremely well suited for studies of its vertical structure. Amongst the asymmetries, a warp in the spine of the on-sky brightness distribution within $\sim80$ au from the star was first unveiled through HST imaging and subsequently confirmed in all follow-up optical/near-IR scattered light observations \citep{Mouillet1997, Heap2000, Golimowski2006, Boccaletti2009, Lagrange2012, Milli2014, Apai2015, Millar-Blanchaer2015}. The presence of this warp was soon attributed to gravitational interaction with an unseen companion in the inner regions of the system \citep{Mouillet1997,Augereau2001}. 
%For simplicity, the physical warp axis has been assumed to lie in the plane of the sky, and to act on a disk that is viewed edge-on; predictions from interaction with a XX$M_{\rm Jup}$ planet located at a position XX-XX AU from the star could explain the observed warp and butterfly asymmetry \citep[e.g.][]{Augereau2001}.
It wasn't until 2009 that a giant planet, $\beta$ Pictoris b, was discovered through direct imaging observations \citep{Lagrange2009, Lagrange2010} at a projected distance of $\sim$9 AU from the star \citep{Lagrange2018b}, misaligned to the outer belt \citep{Lagrange2012} in a way that is roughly consistent with the general expectation from the observed scattered light warp \citep[e.g.][]{Dawson2011, NesvoldKuchner2015}. Interestingly, the position angle of the CO and \ion{C}{1} gas disks is similar to that of the inner disk seen in scattered light, suggesting a link between the two \citep{Matra2017a,Cataldi2018}, but no constraints have been reported so far on the presence of a warp in dust emission at longer wavelengths, due to a lack of angular resolution necessary to resolve this feature.%Since its discovery, various studies have tested whether this planet alone can be held responsible for the observed disk structure. After tentative evidence for the planet being misaligned with the warp, \citet{Dawson2011} showed that this alignment is not needed at present, owing to damping of damping of the planet's orbital inclination due to dynamical friction. Soon after, however, improved follow-up studies showed that the PA of the planet is actually consistent with the position angle of the warp \citep{Lagrange2012}.%In addition, they also showed that if $\beta$ Pic b is aligned with the main disk observed at large radii, no other unseen planet can produce a physical warp that is consistent with the observations.

So far, the observed vertical distribution and width of the belt has received less attention. HST observations found that a sum of two Gaussians or Lorentzians best describes the observed vertical distribution of scattered light emission \citep[e.g.][]{Golimowski2006, Lagrange2012, Apai2015}, with the scale height of the broad Gaussian component reaching values as high as $\sim$12 au at a distance of $\sim$80 au from the star. This corresponds to an aspect ratio of $\sim0.15$ at 80 au, similar to the aspect ratio inferred from modelling of polarimetric near-IR observations \citep[$\sim$0.137,][]{Millar-Blanchaer2015}. 

The vertical distribution of planetesimal belts is a crucial observable as it reflects the inclination distribution of the constituent particles. In our own Kuiper belt, the inclination distribution is broad and bimodal \citep[e.g.][]{Brown2001, Kavelaars2009, Petit2011}, and sets tight constraints on its dynamical history \citep{Morbidelli2008}, particularly on the outward migration of Neptune in the early stages of the Solar System's evolution \citep[e.g.][]{Nesvorny2015}.
Therefore, accessing the inclination distribution of particles tracing planetesimals in extrasolar planetesimal belts is an attractive avenue to probe their dynamics and the effects of unseen planets.

With the assumption of an externally unperturbed and initially cold belt, the inclination distribution (particularly its width) could for example yield a direct link to the relative velocities of planetesimals. This allows us to probe the level of dynamical heating induced by unseen, large gravitational perturbers within the belt \citep[e.g.][]{Quillen2007}. To date, only observations in optical/near-IR scattered light have had sufficient resolution to vertically resolve edge-on belts such as around $\beta$ Pic. Unfortunately, however, the radiation pressure felt by the small grains probed by these observations can significantly excite their eccentricities leading to a consequential increase in their inclinations \citep{Thebault2009}, which has prevented drawing a robust link between the observed vertical structure and the presence of large, unseen bodies required to create it.

In this paper, we tackle this issue using new ALMA interferometric observations that vertically resolve the $\beta$ Pictoris belt at 1.33 mm. The observations are described in \S\ref{sect:obs}, and the main results derived from image analysis in \S\ref{sect:res}. In \S\ref{sect:modelling}, we confirm these results by fitting the interferometric visibilities using parametric models to describe the belt's dust density distribution. In \S\ref{sect:disc}, we link the belt's observed vertical distribution to the inclination distribution of mm grains, and discuss its origin in the context of gravitational perturbations by large bodies within or exterior to the belt. Finally, in \S\ref{sect:concl}, we draw our conclusions and summarize our findings.

\section{Observations}
\label{sect:obs}

\begin{figure*}
%\vspace{-20mm}
 %\centering
 \hspace{-5mm}
   \includegraphics*[scale=0.58]{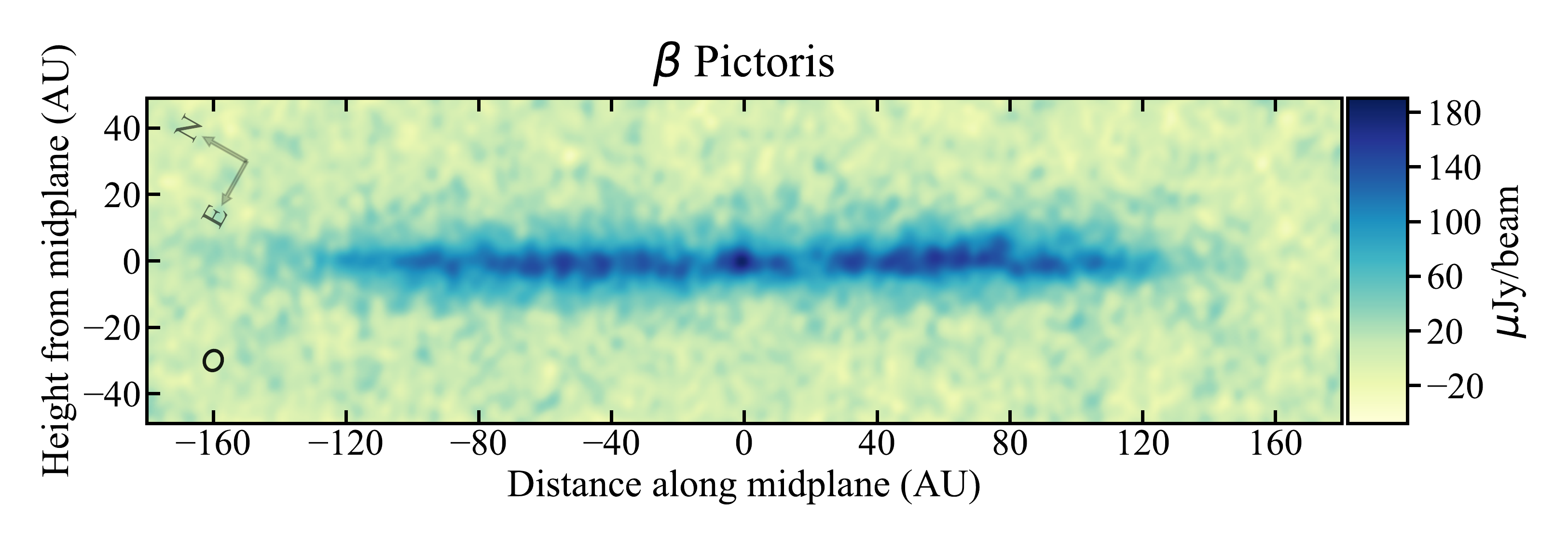}
\vspace{-10mm}
\caption{ALMA 1.33 mm \textsc{CLEAN} images of continuum emission from the $\beta$ Pictoris belt obtained from the combined 12m extended, compact and ACA baselines, with natural weighting. The image is centered at the location of the star (detected here) and was rotated so that the belt midplane aligns with the horizontal axis, assuming the PA of the main disk observed in scattered light (29.3$^{\circ}$). The RMS noise level achieved is 12 $\mu$Jy beam$^{-1}$ for a synthesized beam size of $0\farcs32\times0\farcs27$ ($6.1\times5.3$ au, bottom left circle).%Ciao%. 1D histograms represent probability distributions of each parameter marginalised over the other two, whereas contour maps represent 2D probability distributions of different pairs of parameters, marginalised over the third. Contours represent the central [68.3, 95.5, 99.73] \% of the distribution. Blue solid lines represent marginalised posterior probability distributions of the parameters given our observed sample, and should be compared with the model distributions.
} 
\label{fig:b6contim}
\end{figure*}

%\subsection{ALMA Band 6}
ALMA observations of the $\beta$ Pictoris disk were obtained within its Cycle 2 (project code 2012.1.00142.S) using Band 6 receivers. Data was taken with the 12-m array in extended (one pointing centered on the star) and compact configuration (two pointings centered $\pm5\arcsec$ from the star along the belt midplane), as well as with the Atacama Compact Array (ACA, one pointing centered on the star). Baselines covered a range between 9 and 1574 m, enabling us to be sensitive to structure between about 0$\farcs$3 and 27$\arcsec$. A more complete description of the observations, including calibration, reduction and imaging of CO and other lines can be found in \citet{Matra2017a,Matra2018a}. For continuum imaging, we flagged spectral channels around the frequency of the detected CO J=2-1 line (230.538 GHz), and combined all the datasets from the different ALMA configurations as well as the ACA. In doing so, we applied a constant rescaling factor to the visibility weights of each dataset, where these factors (different for each dataset) were determined from the scatter of the observed visibilities in \S\ref{sect:modmet}. 
Then, we imaged the combined dataset using the CLEAN algorithm \citep{Hogbom1974} in multi-frequency synthesis mode. The dataset covers $\sim$6.8 GHz of total bandwidth at an effective wavelength (frequency) of 1.33 mm (224.58 GHz).
We used natural weighting of the visibilities to ensure maximum sensitivity while achieving a synthesized beam of size $0\farcs32\times0\farcs27$ and PA of -82\fdg01; this implies a resolution of $6.1\times5.3$ au at the distance of the star (19.44 pc).

%{\bf Add a bit more on the observations: pointings of main array and ACA, and adjustment of weights}

%\subsection{Other wavelengths}
%As well as new ALMA continuum observations, we make use of the following datasets:
%\begin{itemize}
%\item HST/Space Telescope Imaging Spectrograph (STIS) optical ($\sim$0.58$\mu$m) scattered light observations, obtained in 2012 and presented in \citet{Apai2015}. The final image has a pixel size of 0\farcs05077 and an approximately twice larger resolution (0\farcs101 or 2.0 AU).
%\item Mid-infrared (mid-IR) dust observations at 11.7$\mu$m and 24.6 $\mu$m obtained with Thermal Region Camera and Spectrograph (T-ReCS) on Gemini South Observatory, obtained between 2003 and 2004 and presented in \citet{Telesco2005}. The images have pixels of size 0\farcs09 and resolution of 0\farcs38 (7.4 AU) and 0\farcs72 (14.0 AU) for the 11.7$\mu$m and 24.6 $\mu$m datasets, respectively.
%\item ALMA CO J=2-1 observations obtained as part of the same observations as the continuum data presented here, and discussed in \citet{Matra2017a}. This data has equal pixel size and approximately equal spatial resolution as the continuum data.
%\end{itemize}
%In this work, we do not reprocess these datasets and simply direct the reader to the relevant works quoted above for a detailed description of the observations and data reduction.

\section{Results}
\label{sect:res}

\subsection{New 1.33 mm dust continuum dataset}

Fig. \ref{fig:b6contim} shows the continuum emission map of the $\beta$ Pictoris system obtained from the combined, naturally weighted, visibility dataset. A peak signal-to-noise ratio (S/N) of 16 is reached at the location of the central star, for the RMS noise level of 12 $\mu$Jy/beam measured in an emission-free region of the continuum map. Emission from the belt is also clearly detected with a peak S/N of $\sim$14, as measured to the SW of the star along the midplane. The spatially integrated 1.33 mm flux measured on the primary-beam-corrected map within a $16\arcsec\times4\arcsec$ box centered on the star is 20$\pm$2 mJy \citep[where the uncertainty is dominated by the 10\% contribution from flux calibration,][]{Fomalont2014}.

The stellar contribution as measured from the peak flux at its location is 190 $\mu$Jy, which should strictly be considered an upper limit due to extra disk emission lying along the line of sight to the star. 
In order to separate the stellar flux from the disk emission, we extract the subset of the dataset containing visibilities at $u-v$ distances $>300 k\lambda$. This $u-v$ cutoff was chosen through inspection of CLEAN images, as an optimal trade-off to both ensure that all large-scale emission from the disk is filtered out, while retaining as much data as possible for maximum sensitivity. The obtained CLEAN image has an RMS noise level of 18 $\mu$Jy/beam for a beam size of $0\farcs22\times0\farcs19$, and a compact point source is detected at the stellar position. A simple 2D Gaussian fit within CASA yields a stellar flux of 80$\pm$20 $\mu$Jy at 1.33 mm.
This detection is consistent with the expected stellar flux of $\sim$86 $\mu$Jy obtained by extrapolating the Rayleigh-Jeans tail of the stellar blackbody-like emission from near-IR wavelengths. We therefore assume a stellar flux of 86 $\mu$Jy at 1.33 mm throughout the rest of this paper. 

\begin{figure}
%\vspace{-20mm}
 %\centering
 \hspace{-5mm}
  \includegraphics*[scale=0.33]{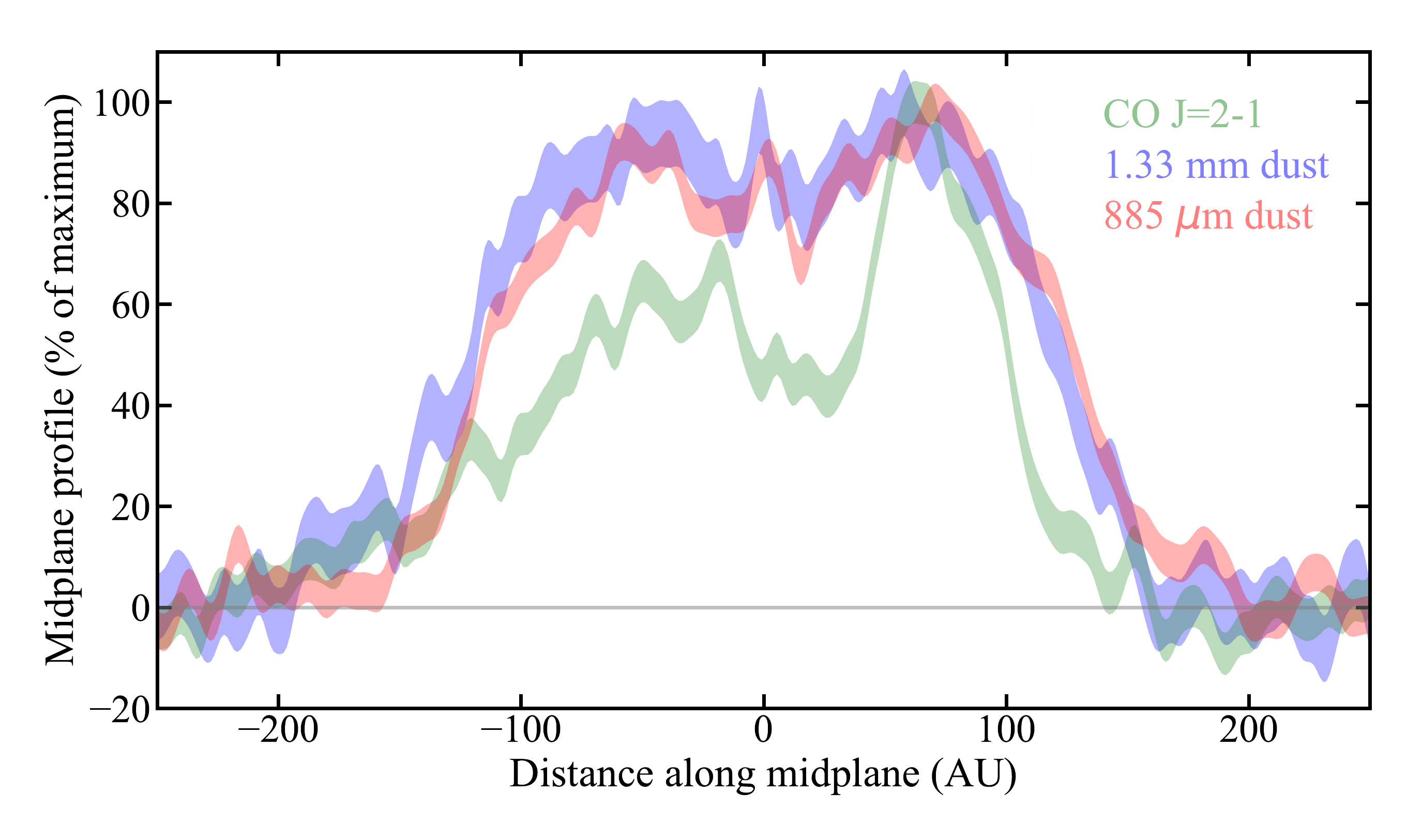}
\vspace{-7mm}
\caption{Radial distribution of continuum 1.33 mm (blue), 885 $\mu$m (red) and CO J=2-1 emission (green) obtained by spatially integrating images (such as Fig. \ref{fig:b6contim}) within a height of $\pm20$ au from the belt midplane. Each profile is normalised to its maximum to enable direct comparison. %. 1D histograms represent probability distributions of each parameter marginalised over the other two, whereas contour maps represent 2D probability distributions of different pairs of parameters, marginalised over the third. Contours represent the central [68.3, 95.5, 99.73] \% of the distribution. Blue solid lines represent marginalised posterior probability distributions of the parameters given our observed sample, and should be compared with the model distributions.
} 
\label{fig:radprof}
\end{figure}

\subsection{Belt radial structure}

In Fig. \ref{fig:radprof}, we show the profile of the continuum emission vertically integrated within $\pm20$ au of the disk midplane, defined here as the position angle of the main disk observed in scattered light \citep[29.3$^{\circ}$,][]{Lagrange2012}. Contrary to what is seen for CO emission in the same dataset (green line), the millimetre-sized dust traced by the 1.33 mm continuum observations (blue line) shows no significant peak brightness asymmetry between the SW and NE sides of the disk. This is in contrast with the SW/NE brightness asymmetry, between 30 and 80 au on either side the star, reported at 885 $\mu$m by \citet{Dent2014}.
%the previous result reported by \citet{Dent2014}, and confirmed by our reanalysis of the ALMA 885$\mu$m continuum dataset (red line), that there is a SW/NE brightness asymmetry between 30 and 80 au from the star. 
The SW/NE ratio measured within the same radial distances as \citet{Dent2014}, vertically integrated between $\pm20$ au, is 1.03$\pm$0.04 in the new 1.33 mm dataset, consistent with no asymmetry. The same SW/NE ratio is 1.05$\pm$0.03 in the 885 $\mu$m archival dataset, indicating again a negligible (5$\pm$3)\% asymmetry in this radial region. As this measurement can be affected by imaging in the presence of incomplete $u-v$ sampling and subsequent deconvolution via the CLEAN algorithm, we postpone further discussion of this issue in the context of visibility models to \S\ref{sect:simpledisk_dbgauss_band7}. %A radial size asymmetry similar to the scattered light \citep[e.g.][]{KalasJewitt1995} is also observed, for both CO and 1.33 mm dust, with emission on the NE side detected out to larger radii ($\sim$200 AU) than the SW side ($\sim$160 au). %This is not significantly detected in the 885 $\mu$m dataset, which may be caused by the effect of interferometric spatial filtering of large scale, smooth emission as seen on the NE side of the belt. Although this effect is hard to quantify, it is likely affecting the 885 $\mu$m dataset given that its shortest baseline (15m) only probes emission on scales up to $\sim$12.2$\arcsec$, which is smaller than the on-sky extent of the disk ($\sim$16.5$\arcsec$ for radii up to 160 au).

%{\bf Where does 12.2 arcsec come from?}

\subsection{Belt vertical structure}

\begin{figure*}
%\vspace{-20mm}
 %\centering
 \hspace{-5mm}
  \includegraphics*[scale=0.36]{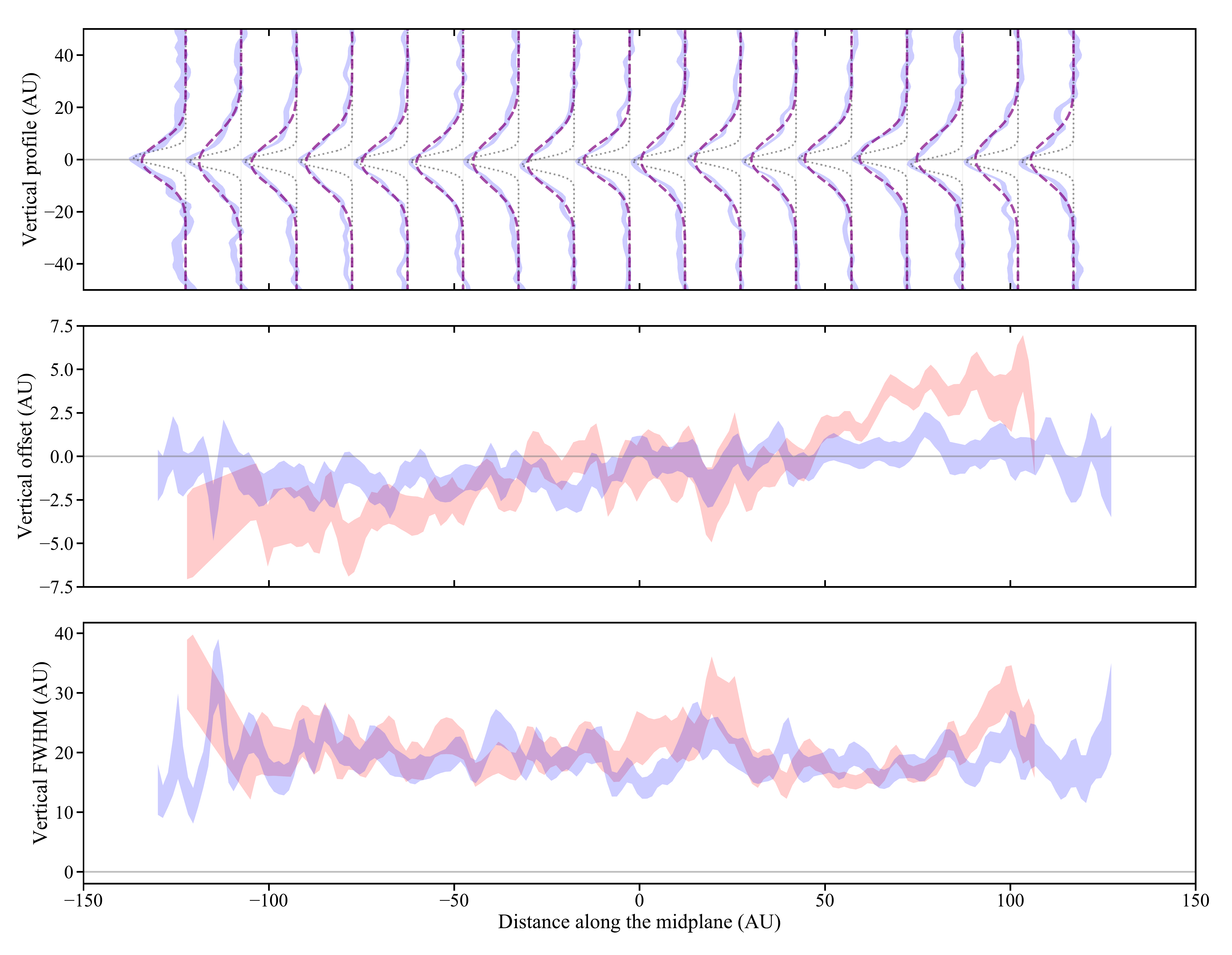}
\vspace{-5mm}
\caption{Vertical structure of continuum 1.33 mm emission (blue) and CO J=2-1 emission (red) from the belt. Shaded regions indicate $\pm1\sigma$ uncertainty ranges. \textit{Top:} Blue shaded regions represent normalised vertical emission profiles measured at midplane locations shown by the x axis (see main text for details). Dotted lines represent the resolution element of the observations, whereas purple dashed lines are single-Gaussian fits to each of the observed profiles. \textit{Middle:} Shaded regions show the disk spines, i.e. the centroids of the vertical Gaussians fitted in the top panel, as a function of radius. CO J=2-1 emission (red) is considerably misaligned with respect to 1.33mm emission (blue) from the same ALMA dataset. \textit{Bottom:} Vertical widths (FWHM) of the Gaussians fitted in the top panel, after deconvolving by the FWHM of the beam of our observations (projected on the perpendicular to the disk's midplane). These show similar heights for both CO and continuum emission.}   %. 1D histograms represent probability distributions of each parameter marginalised over the other two, whereas contour maps represent 2D probability distributions of different pairs of parameters, marginalised over the third. Contours represent the central [68.3, 95.5, 99.73] \% of the distribution. Blue solid lines represent marginalised posterior probability distributions of the parameters given our observed sample, and should be compared with the model distributions.

\label{fig:vertcombo}
\end{figure*}

\begin{figure}
%\vspace{-20mm}
 %\centering
 \hspace{-4mm}
  \includegraphics*[scale=0.36]{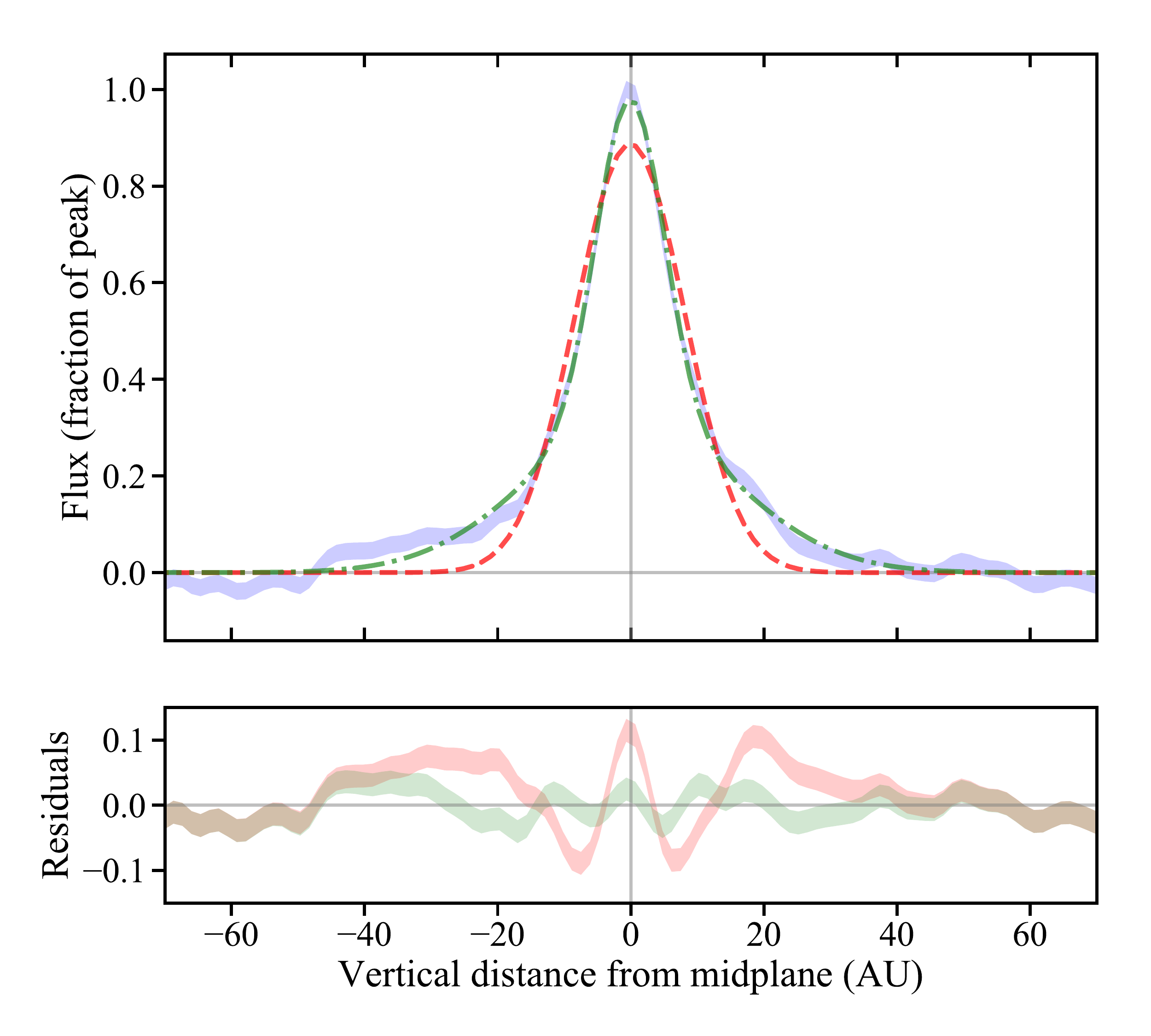}
\vspace{-8mm}
\caption{\textit{Top:} Vertical profile integrated along the belt's sky-projected midplane of 1.3 mm emission (blue shaded region) as a function of height from the midplane. A best-fit Gaussian model is shown in red, whereas a best-fit double Gaussian model is shown in green. \textit{Bottom:} Residuals obtained after subtracting the best-fit Gaussian (red) and double Gaussian (green) models from the data, showing that the double Gaussian model is significantly better at reproducing the observed vertical profile.
} 
\label{fig:vertprofs_radint}
\end{figure}

Fig. \ref{fig:vertcombo} (top) shows vertical profiles of emission perpendicular to the disk midplane as a function of distance to the star, which were obtained by integrating emission every 15 au along the radial (midplane) direction. As shown by comparing the profiles (blue shaded region) to the resolution element of our observation (black dotted line), the dust disk is clearly resolved in the vertical direction.
Following the procedure described in detail in \S3.2 and Appendix B of \citet{Matra2017a}, we fit Gaussians (purple dashed lines) to the vertical profiles at each midplane location, and derive vertical best-fit Gaussian centroids (forming the disk spine, Fig. \ref{fig:vertcombo} middle panel) and widths (measured as FWHM, Fig. \ref{fig:vertcombo} bottom panel). We report two main findings:

\begin{enumerate}
\item In contrast with the CO J=2-1 line emission, we find that dust emission is not well represented by a single vertical Gaussian, being more centrally peaked and having broader wings extending out to $\pm$40 au from the sky-projected midplane. This is evident in Fig. \ref{fig:vertprofs_radint}, which shows the residual vertical profile obtained by integrating emission across the entire midplane after vertically centering it so that the centroid of the emission aligns with the midplane. The best-fit Gaussian profile (red, with a standard deviation of $8.2\pm0.2$ au) leaves significant residuals (Fig. \ref{fig:vertprofs_radint}, bottom), whereas a double-Gaussian profile (green, with standard deviations of $5.1\pm0.3$ and $15.7\pm1.4$ au) considerably improves the fit, leaving no significant residual emission. We find that relative flux calibration systematics of $\pm10\%$ between the datasets with compact and extended baselines (tracing vertically broader and narrower emission, respectively) cannot account for this discrepancy, which must therefore be physical in origin. This shows that the double-Gaussian or Lorentzian profiles needed to fit scattered light observations \citep{Golimowski2006, Lagrange2012, Apai2015} reflect the vertical structure of the millimeter emission that traces the dust-producing planetesimals.

\item The belt spine (Fig. \ref{fig:vertcombo}, middle) presents little vertical displacement along the midplane and is largely consistent with the PA of the main disk observed in scattered light. This is significantly different from the CO J=2-1 emission, whose spine presents an extra tilt angle dPA of $\sim$4$^{\circ}$ similar to that of the warp interior to $\sim$80 au seen in scattered light \citep{Apai2015,Matra2017a}. Therefore, we conclude that an inner disk tilt as large as observed in scattered light and CO observations is not detected in the structure of millimetre grains. Assuming the mm grains are tracing the planetesimal distribution, this shows that planetesimals are not perturbed into a warp - or at least not into one as pronounced as that observed in the distribution of the smallest grains observed in scattered light. At the same time, the tilt present in both the scattered light and the CO distribution, with the CO clump clearly offset by 4-6 au above the main disk midplane, suggests a connection between the two.

%Peaks are 124+-8 and 56+-9 for double and 163+-3 for single.

\end{enumerate}

%Assuming that mm grains are good tracers of the planetesimal distribution in the disk, th

%The rightmost profile is obtained by summing over all radii on both sides of the star, after centering each vertical profile 

\section{Modelling}
\label{sect:modelling}

In this section, we aim to confirm our results inferred from images through radiative transfer modelling of the interferometric visibility datasets, enabling us to constrain the 3D distribution of mm dust in the belt while avoiding potential artifacts introduced by the imaging and CLEAN deconvolution processes.

\subsection{Method}
\label{sect:modmet}
We model the $\beta$ Pic belt using mm dust density distributions described in the following subsections, and the star as a point source with a flux density of 86 $\mu$Jy located in the geometrical center of the circular belt. For each model realization, we solve the radiative transfer at 1.33mm using RADMC-3D\footnote{\url{http://www.ita.uni-heidelberg.de/~dullemond/software/radmc-3d/}}, and assuming the dust radial temperature profile expected from a blackbody around a star of 8.7 L$_{\odot}$ \citep{KennedyWyatt2014}. Since we are interested in the structure rather than the overall amount of material in the belt, we fix the dust opacity and fit for the belt mass as a normalization factor, noting that this does not affect our conclusions on the belt structure. In addition to the parameters that govern the belt dust density distribution and its overall mass, we fit for belt geometry parameters (inclination $I$ and position angle PA).

Having created the model belt image, we produce visibility data separately for each of the ALMA arrays (12m and ACA), for each configuration of the 12m array and for each of the two mosaic pointings used for the compact configuration observations. We do so by first shifting the model belt image by RA and Dec offsets that we leave as free parameters in the fit, and multiplying it by the primary beam response of the relevant array, at its appropriate sky location for each observation. We then Fourier transform the image to produce a grid of model complex visibilities, which we evaluate at the $u-v$ locations sampled by the observation in question as described e.g. in \citet{Marino2016} and \citet{Tazzari2018}. Finally, we add a point source representing the star to the model visibilities, located in the geometric center of the belt, and with its flux appropriately attenuated given its location with respect to the primary beam center.

The model visibilities are then fitted to the observed visibilities through a Monte Carlo Markov Chain (MCMC) method implemented through the \textsc{emcee} package \citep{Foreman-Mackey2013}. Eight nuisance parameters for the spatial shifts (in RA and Dec, so two for each dataset), two parameters to determine the viewing geometry ($I$ and PA), one normalization parameter setting the mass of the belt for a fixed opacity, and n parameters governing the dust distribution make up the 11+n - dimensional parameter space that we explore in our fit. Through the \textsc{emcee} package, we sample the 11+n dimensional posterior probability distribution of the explored parameters using the affine-invariant MCMC ensemble sampler of \citet{GoodmanWeare2010}. The posterior probability distribution is computed assuming uniform priors on all of the probed parameters, and a likelihood function proportional to $e^{-\chi^2/2}$, where $\chi^2$ is the usual definition of the chi-square function. The MCMC was run for each tested model with a number of walkers equal to 10x the number of free parameters, and over 2500 steps, ensuring each of the chains had reached convergence.

The uncertainty $\sigma$ on each visibility datapoint is defined using the visibility weights $w$ delivered by the ALMA observatory, where $\sigma_{u_i-v_i}=1/\sqrt{w_{u_i-v_i}}$. While assuming the delivered weights and hence the uncertainties are correct for different $u-v$ points relative to each other, we left a constant rescaling factor, equal for all $u-v$ points within a given dataset, as a free parameter in our model fit of \S \ref{sect:simpledisk}. This is justified by previous ALMA modelling studies also needing to rescale the weights to appropriately represent the true scatter in the visibilities \citep[e.g.][]{Marino2018b}. We find this rescaling factor to be different but very well constrained to $0.4508\pm0.0006$, $0.927\pm0.014$, $0.690\pm0.003$, and $0.698\pm0.003$ for the 12m extended, ACA and for the two pointings of the 12m compact configuration, respectively. We therefore fix the rescaling factors to these values before producing the image in Fig. \ref{fig:b6contim}, and for all subsequent model fitting runs.

\begin{deluxetable*}{ccccccc}
\tabletypesize{\scriptsize}
\tablecaption{Best-fit model parameters \label{tab:modelcomparison}}
\tablewidth{0pt}
\tablehead{
\colhead{Parameter} & \colhead{Unit} & \colhead{Model 1} & \colhead{Model 2} & \colhead{Model 3}  & \colhead{Model 4}  & \colhead{Model 5} % & \colhead{$\Delta\chi^2$} & \colhead{$\Delta$ BIC} &
%\colhead{$\Delta$ AIC} %\\
%\colhead{ } & \colhead{ } & \colhead{(GHz)} & \colhead{(Jy km/s)} & \colhead{ } &
%\colhead{ }
}
\startdata
$r_{\rm c}$ & au & $104.9_{-1.1}^{+1.1}$ & $107.5_{-1.0}^{+1.0}$ & $106.3_{-1.3}^{+1.4}$ & $107.8_{-1.0}^{+1.0}$ & $105.2_{-1.4}^{+1.4}$ \\
$\sigma$ & au & $38.9_{-1.1}^{+1.3}$ & $36.7_{-1.3}^{+1.0}$ & $36.6_{-1.2}^{+1.1}$ & $36.4_{-1.3}^{+1.1}$ & $36.3_{-1.2}^{+1.1}$ \\
$h$ &  & $0.070_{-0.004}^{+0.004}$ & $0.014_{-0.006}^{+0.006}$ & $^{\ast}0.020_{-0.008}^{+0.010}$ & $0.054_{-0.004}^{+0.004}$ & $^{\ast}0.072_{-0.012}^{+0.014}$ \\
$I$ & $^{\circ}$ & $86.6_{-0.3}^{+0.4}$ & $88.6_{-0.3}^{+0.2}$ & $88.8_{-0.3}^{+0.4}$ & $89.3_{-0.5}^{+0.4}$ & $89.5_{-0.4}^{+0.3}$ \\
$PA$ & $^{\circ}$ & $30.0_{-0.1}^{+0.1}$ & $29.7_{-0.1}^{+0.1}$ & $29.7_{-0.1}^{+0.1}$ & $29.8_{-0.1}^{+0.1}$ & $29.7_{-0.1}^{+0.1}$ \\
$h_2$ &  & - & $0.110_{-0.006}^{+0.008}$ & $^{\ast}0.14_{-0.02}^{+0.02}$ & - & - \\
$a$ &  & - & $0.20_{-0.04}^{+0.05}$ & $0.21_{-0.04}^{+0.05}$ & - & - \\
$\beta$ &  & - & - & $0.7_{-0.2}^{+0.2}$ & - & $0.4_{-0.2}^{+0.2}$ \\
$p$ &  & - & - & - & $0.9_{-0.1}^{+0.1}$ & $0.85_{-0.07}^{+0.07}$  \\
%$r_{\rm c}$ & au & $104.9_{-1.1}^{+1.1}$ & $107.5_{-1.0}^{+1.0}$ & $106.3_{-1.3}^{+1.4}$ \\
%$\sigma$ & au & $38.9_{-1.1}^{+1.3}$ & $36.7_{-1.3}^{+1.0}$ & $36.6_{-1.2}^{+1.1}$ \\
%$H$ & au & $3.5_{-0.2}^{+0.2}$ & $0.7_{-0.3}^{+0.3}$ & $1.0_{-0.4}^{+0.5}$ \\
%$i$ & $^{\circ}$ & $86.6_{-0.3}^{+0.4}$ & $88.6_{-0.3}^{+0.2}$ & $88.8_{-0.3}^{+0.4}$ \\
%$PA$ & $^{\circ}$ & $30.0_{-0.1}^{+0.1}$ & $29.7_{-0.1}^{+0.1}$ & $29.7_{-0.1}^{+0.1}$ \\
%$H_2$ & au & - & $5.5_{-0.3}^{+0.4}$ & $6.8_{-1.0}^{+1.3}$ \\
%$a$ &  & - & $0.20_{-0.04}^{+0.05}$ & $0.21_{-0.04}^{+0.05}$ \\
%$\beta$ &  & - & - & $0.7_{-0.2}^{+0.2}$ \\
%$r_{\rm c}$ & au & $102.7_{-1.4}^{+1.4}$ & $103.1_{-1.0}^{+1.0}$ & $106.3_{-1.3}^{+1.4}$ \\
%$\sigma$ & au & $36.5_{-1.6}^{+1.7}$ & $36.7_{-1.1}^{+1.2}$ & $36.6_{-1.2}^{+1.1}$ \\
%$H$ & au & $7.9_{-0.3}^{+0.3}$ & $2.3_{-0.7}^{+0.6}$ & $1.0_{-0.4}^{+0.5}$ \\
%$i$ & $^{\circ}$ & $>87.0$ & $89.3_{-0.4}^{+0.4}$ & $88.8_{-0.3}^{+0.4}$ \\
%$PA$ & $^{\circ}$ & $29.9_{-0.1}^{+0.2}$ & $29.7_{-0.1}^{+0.1}$ & $29.7_{-0.1}^{+0.1}$ \\
%$H_2$ & au & - & $11.3_{-0.6}^{+0.8}$ & $6.8_{-1.0}^{+1.3}$ \\
%$a$ &  & - & $0.23_{-0.04}^{+0.05}$ & $0.21_{-0.04}^{+0.05}$ \\
%$\beta$ &  & - & - & $0.7_{-0.2}^{+0.2}$ \\
\hline
Free Parameters & & 14 & 16 & 17 & 15 & 16 \\
$\chi^2$ & & 1456364.9 & 1456297.6 & 1456297.4 & 1456312.2 & 1456297.3\\
$\Delta\chi^2$ & & - & -67.3\tablenotemark{a} & -0.2\tablenotemark{b} & +14.6\tablenotemark{b} & -0.3\tablenotemark{b} \\
$\Delta$AIC & & - & -63.3\tablenotemark{a} & +1.8\tablenotemark{b} & +12.6\tablenotemark{b} & -0.3\tablenotemark{b} \\
$\Delta$BIC & & - & -38.9\tablenotemark{a} & +14.0\tablenotemark{b} & +0.33\tablenotemark{b} & -0.3\tablenotemark{b} \\
%$\chi^2$ & & 1456439.7 & 1456311.3 & 1456297.4\\
%$\Delta\chi^2$ & & - & -128.4 & -13.9 \\
%$\Delta$AIC & & - & -124.4 & -11.9 \\
%$\Delta$BIC & & - & -100.0 & -0.3 \\
\enddata
\label{table:bestfit}
%% Text for table notes should follow after the \enddata but before
%% the \end{deluxetable}. Make sure there is at least one \tablenotemark
%% in the table for each \tablenotetext.

%\tablecomments{Table \ref{tab:mols} is published in its entirety in the electronic edition of the {\it Astrophysical Journal}. A portion is shown here for guidance regarding its form and content.}
\tablecomments{\textit{Model 1}: Radially Gaussian ring with peak $r_c$ and standard deviation $\sigma$, with a single-component, Gaussian vertical profile with a radially constant aspect ratio $h$. \textit{Model 2}: Same as model 1, with a second, broader Gaussian component to the vertical profile, with aspect ratio $h_2$. The ratio of the peak value of the second component to that of the first component is $a$. \textit{Model 3}: Same as model 2, but with a radially varying scale height for both Gaussian components, following $H\propto r^{\beta}$, or equivalently $h\propto r^{\beta-1}$. \textit{Model 4}: Same as model 1, but with a vertical profile described by a non-Gaussian functional form (where $\rho\propto exp\left[-\left(\frac{\lvert z\rvert}{\sqrt{2}hr}\right)^{p}\right]$, see Eq. 3), with a radially constant aspect ratio $h$. \textit{Model 5}: Same as model 4, but with a radially varying aspect ratio $h\propto r^{\beta-1}$. For Model 3 and 5, $^{\ast}$ symbols indicate that aspect ratios are quoted at a radius of 50 au.} %References: (1) This work; (2) \citet{Matra2017a}; (3) \citet{Dent2014}
\tablenotetext{a}{Difference with respect to Model 1}
\tablenotetext{b}{Difference with respect to Model 2}
\end{deluxetable*}

\subsection{Radially and vertically Gaussian, axisymmetric model with radially constant aspect ratio}
\label{sect:simpledisk}
%Motivated by the general morphology of the radial and vertical profiles of the belt in \S\ref{sect:res}, 
We begin by assuming the simplest dust density distribution model with the minimum number of free parameters, namely a radially and vertically Gaussian, axisymmetric belt with radially constant aspect ratio. The distribution of the dust density $\rho$ in the belt reads:
\begin{equation}
\rho=\rho_0\ e^{-\frac{(r-r_{\rm c})^2}{2\sigma^2}}\frac{e^{-\frac{z^2}{2(hr)^2}}}{\sqrt{2\pi}hr}
\end{equation}
where $r$ and $z$ are cylindrical coordinates, $r_{\rm c}$ and $\sigma$ are the center and standard deviation of the radially Gaussian belt, $h$ is the aspect ratio (here independent of radius), and $\rho_0$ is a normalization factor proportional to the total dust mass in the belt, which we fit for.

%{\bf The $\rho_0$ normalization in the two equations is not exactly consistent because of the vertical structure prefactors, which is a bit confusing. Should I just add a normalization to the first Gaussian? Check the code.}

The best-fit parameters of this model, defined as the $50^{+34}_{-34}$th percentiles of their marginalized posterior probability distributions, are listed in Table \ref{tab:modelcomparison} (Model 1). Within the framework of this model, and in qualitative agreement with Fig. \ref{fig:b6contim}, we find the disk to be near edge-on, with an inclination from face-on of $86.6^{+0.4}_{-0.3}$ degrees, and with its Gaussian radial distribution being centred at $\sim$105 AU with a width of $\sim$92 AU.

Fig. \ref{fig:combores}, top left shows the residuals obtained after subtracting the best-fit model visibilities from the data, and imaging the residual visibility dataset. The bottom left panel of Fig. \ref{fig:combores} shows the residual radial profile constructed as in Fig. \ref{fig:radprof}; the lack of significant residuals and SW/NE asymmetries reaffirms our conclusion that no clear departure from axisymmetry is present, within our measurement uncertainties, in the ALMA 1.33 mm dataset.

On the other hand, the right panel shows the vertical profile obtained after radially integrating the belt's residual emission as done in Fig. \ref{fig:vertprofs_radint}. This displays significant residuals very similar to those obtained by simply subtracting a single Gaussian from the vertical flux distribution in the image plane. This confirms that the vertical distribution of mm dust in the planetesimal belt around $\beta$ Pictoris cannot be reproduced by a single Gaussian.

\subsection{Double-Gaussian vertical structure model, with radially constant aspect ratio}
\label{sect:simpledisk_dbgauss}

\begin{figure*}
%\vspace{-20mm}
 %\centering
 \hspace{-14mm}
  \includegraphics*[scale=0.39]{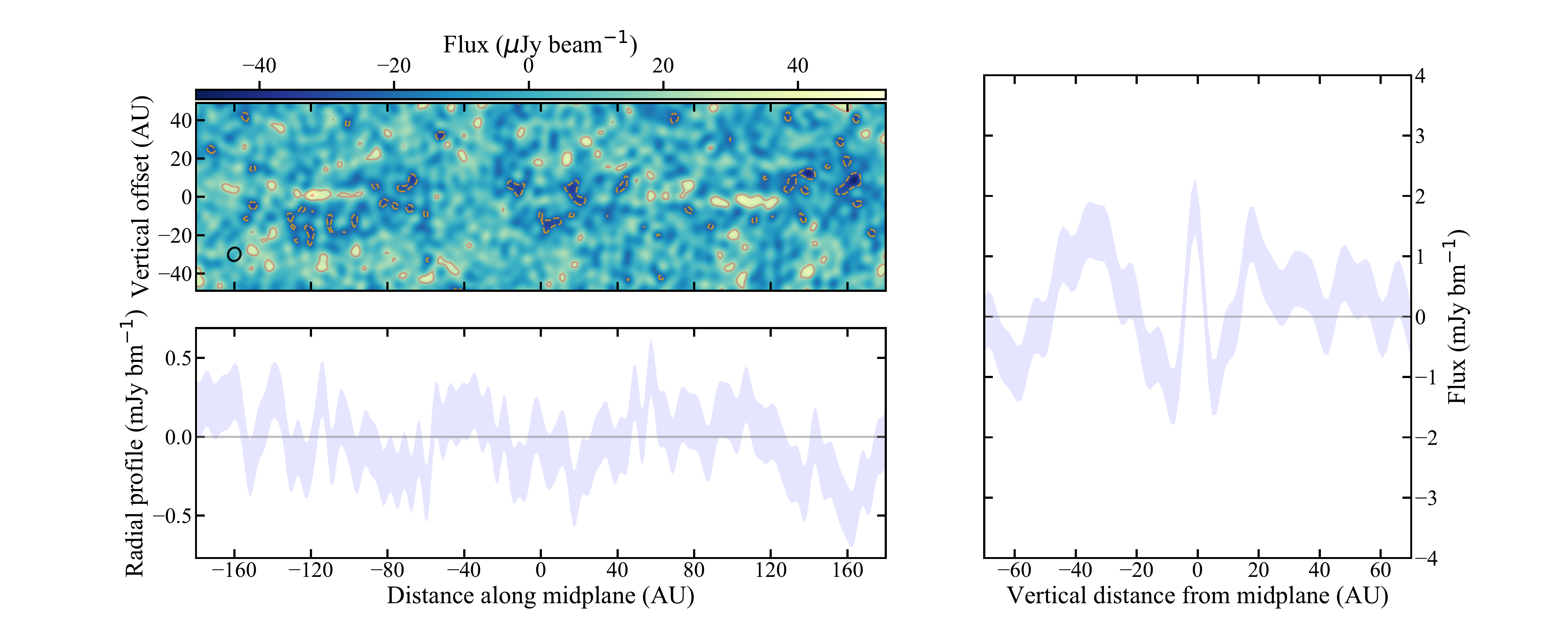}
%\vspace{-28mm}
\caption{\textit{Top left:} Dirty image of the residual visibilities after subtracting the best-fit axisymmetric model with a vertical Gaussian density distribution (Model 1, see \S\ref{sect:simpledisk} and Table \ref{table:bestfit} for best-fit parameters) from the 1.33 mm data. Contours represent $\pm$2,4,6.. times the RMS noise level. \textit{Bottom left:} Radial profile obtained from the residual image by vertically integrating within $\pm20$ au of the midplane, as in Fig.\ref{fig:radprof}, showing no significant SW/NE asymmetry. \textit{Right:} Radially averaged vertical profile of the residual image, showing significant residuals analogous to the simple fit of Fig. \ref{fig:vertprofs_radint}. These indicate that a single vertical Gaussian distribution is not a good fit to the data. 
} 
\label{fig:combores}
\end{figure*}

\begin{figure*}
\vspace{-5mm}
 %\centering
 \hspace{-14mm}
  \includegraphics*[scale=0.39]{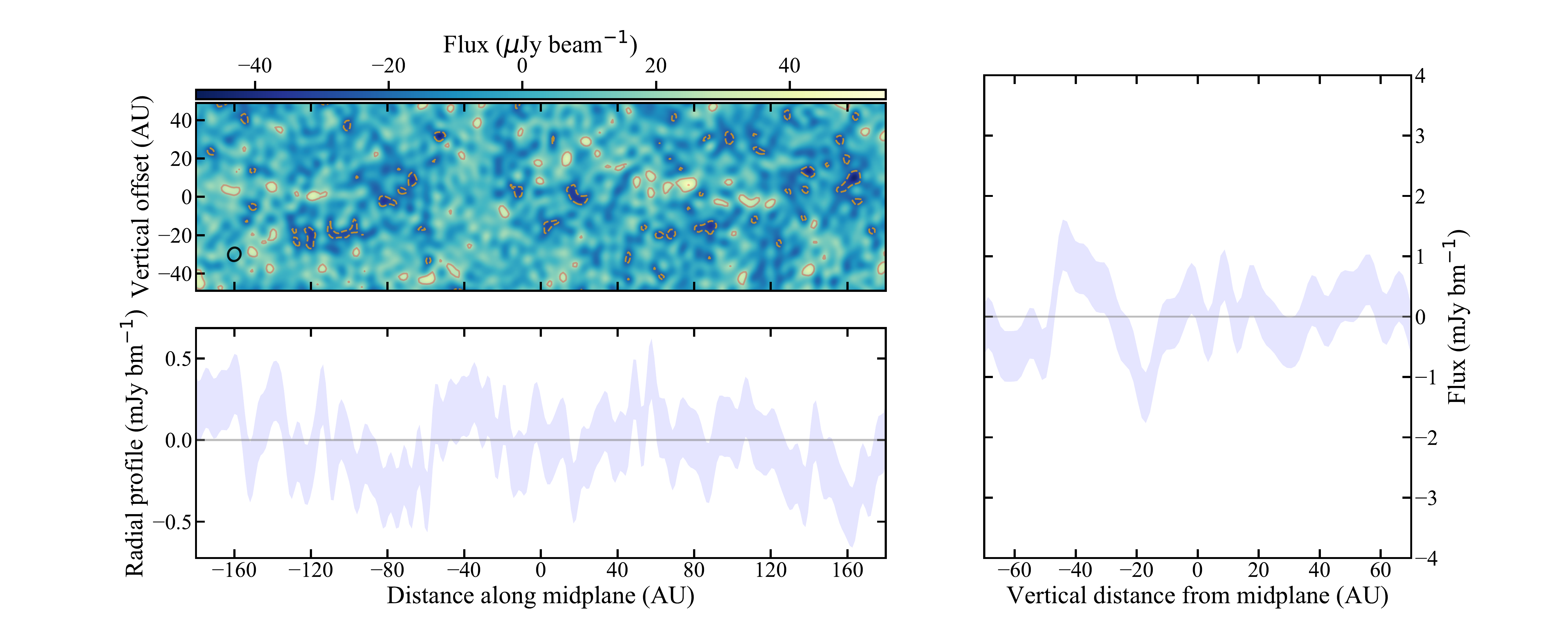}
%\vspace{-28mm}
\caption{Same as \ref{fig:combores}, but for a model where the vertical dust density distribution is the sum of two Gaussians (Model 2, see \S\ref{sect:simpledisk_dbgauss} and Table \ref{table:bestfit} for best-fit parameters). As demonstrated by the vertical profile of the residuals (right), this model produces a much better fit to the observations.
} 
\label{fig:combores_dbgauss}
\end{figure*}

Motivated by the appearance of the vertical emission distribution in the image plane, we construct a belt model with a vertical distribution characterised by a narrower and broader Gaussian. The Gaussians are both centered in the belt's midplane ($z=0$) and are characterized, respectively, by aspect ratios $h$ and $h_2$ leading to standard deviations (scale heights) $hr$ and $h_2r$ at any given radius $r$. They comprise a fraction $a$ and $1-a$ of the total belt mass.
The functional form of the dust density distribution thus reads
\begin{equation}
\rho=\rho_0\ e^{-\frac{(r-r_{\rm c})^2}{2\sigma^2}}\left[\frac{a}{\sqrt{2\pi}hr}e^{-\frac{z^2}{2(hr)^2}}+\frac{1-a}{\sqrt{2\pi}h_2r}e^{-\frac{z^2}{2(h_2r)^2}}\right].
\label{eq:2}
\end{equation}
This model is significantly better at describing the vertical distribution of the belt, as seen when comparing its residuals (Fig. \ref{fig:combores_dbgauss}) to the single vertical Gaussian model (Fig. \ref{fig:combores}). Comparing best-fit models (Model 2 and 1, respectively, in Table \ref{tab:modelcomparison}) through an Akaike or Bayesian Information Criterion (AIC, or BIC), which penalise models with a larger number of free parameters, clearly favors the model with the more complex vertical structure. Formally, the large BIC differential (or analogously the ratio of the model likelihoods) indicates that model with a double Gaussian vertical structure is $>10^{8}$ times more likely to be better at describing the data than the model with a single Gaussian, providing strong evidence in support of our conclusion.

The narrow Gaussian component has a vertical FWHM of $\sim$2 au at 150 au, which is similar to the size of our resolution element and only resolved through the forward modelling presented here. Note also that the inclination of the double Gaussian model is significantly higher (closer to edge-on) than found for the single Gaussian model. This is likely attributable to the single Gaussian model being driven to a less edge-on configuration in an attempt to capture the broad sky-projected vertical distribution of the disk emission. The position angle East of North of the mm dust disk ($29.7_{-0.1}^{+0.1}$) is marginally consistent within the uncertainties with the outer disk observed in scattered light \citep[$29.3_{-0.1}^{+0.2}$,][]{Lagrange2012}.
%{\bf Maybe comment on the inclinations in Model 1 vs. Model 2/3?}

\subsection{Adding model complexity:\\ radially varying aspect ratio}
\label{sect:simpledisk_dbgauss_withbeta}

The appearance of the residuals in Fig. \ref{fig:combores_dbgauss}, after subtracting a model where the aspect ratio is the same at all belt radii, indicates that we have already achieved a good fit to the data. Nonetheless, we test whether we can detect a radial dependence of the vertical aspect ratio through a model where the scale height is a function of radius, and this dependence is parametrized through a flaring parameter $\beta$. The dust density distribution reads the same as in Eq. \ref{eq:2}, but with $h$ and $h_2$ being a function of radius following $h(r)=h_{\rm50 au}\left(\frac{r}{\rm 50 au}\right)^{\beta-1}$ and similarly $h_2(r)=h_{2, \rm50 au}\left(\frac{r}{\rm 50 au}\right)^{\beta-1}$. Then, $\beta=1$ indicates a constant aspect ratio $h=H(r)/r$ as considered in the previous models, $\beta>1$ a flared disk, and $\beta=0$ a disk with a radially constant scale height.

The marginalized posterior probability distribution of $\beta$ results in $\beta=0.7^{+0.2}_{-0.2}$, indicating that models with a scale height increasing with radius ($\beta>0$) are significantly preferred, with a marginal improvement achieved for models with an aspect ratio decreasing with radius ($\beta<1$). However, %we then employ a BIC test to compare the formal best-fit for this model (Model 3 in Table \ref{tab:modelcomparison}) with that for the previous model runs with a radially constant scale height (Model 2). 
we find no evidence that the extra free parameter $\beta$ leads to an overall improvement in the fit to the data, which is confirmed by visual inspection of the residuals.
We therefore conclude that our data supports models where the scale height increases with radius, with a flaring parameter $\beta$ consistent with one.% However, the data does not contain sufficient information to robustly discern between models with $\beta$ values different .
%
%{\bf My take on the above is that fitting for $\beta$ gives a result close to one, in support of the simple constant aspect ratio model. Maybe say something about the physical plausibility of this?}

\subsection{Functional forms other than Gaussians}
\label{sect:simpledisk_withp}

Our results do not rule out that a different parametric prescription for the vertical density distribution could describe the data as well as the double Gaussian parametrization adopted here. For example, we here attempt to use a model where the dropoff of the vertical density distribution differs from a Gaussian \citep[as implemented by][]{Ahmic2009}. The functional form reads
\begin{equation}
\rho=\rho_0\ e^{-\frac{(r-r_{\rm c})^2}{2\sigma^2}}Ce^{-\left(\frac{\lvert z\rvert}{\sqrt{2}hr}\right)^{p}},
\end{equation}
with $h(r)=h_{\rm50 au}\left(\frac{r}{\rm 50 au}\right)^{\beta-1}$ (as in \S\ref{sect:simpledisk_dbgauss_withbeta}), and $C=\left[\int^{+\infty}_{-\infty}{e^{-\left(\frac{\lvert z\rvert}{\sqrt{2}hr}\right)^{p}}}\right]^{-1}$ ensuring normalization of the vertical distribution. Distributions with $p<2$ allow the vertical distribution to drop faster than a Gaussian the further we move from the midplane, but to have broader wings; \citet{Ahmic2009} found this scenario to be preferred in scattered light observations of the $\beta$ Pic belt, with $p$ values around or below 1.

We tried two model runs with the above functional form and $p$ left as a free parameter. In the first (Model 4), we fix the aspect ratio ($\beta=1$) as done for the Model 1 and 2 runs. In the second run (Model 5), we also leave the flaring parameter $\beta$ free. Our fitting results for both model runs, shown in Table \ref{tab:modelcomparison}, indicate that $p\sim0.9$, in broad agreement with the scattered light findings, and consistent with the broad-winged distribution derived from the image analysis.

We also find that best-fit Model 4 and 5 (with $p=0.9$) are significantly better than a single Gaussian (Model 1). Compared with the double Gaussian, constant aspect ratio model (Model 2, see Table \ref{tab:modelcomparison}, bottom), we find that Model 4 (with $\beta$ fixed to 1) is slightly worse, whereas Model 5 (with $\beta\sim0.4$) is just as good at reproducing the data, with extremely similar residuals.
In summary, there is no evidence in the mm data that a model with a single population of mm grains that has a vertical dropoff rate that is shallower than Gaussian (Eq. 3 with $p\sim0.9$, Models 4 and 5) is preferred to a double Gaussian model (Model 2). In fact, the convergence of Models 4 and 5 to a broad-winged distribution confirms our main finding that there exists a population of high inclination (dynamically hot) particles in the $\beta$ Pic disk. For later interpretation, we prefer to use the double Gaussian simply to facilitate direct comparison with the Kuiper belt (see \S\ref{sect:KBcomp}).

%We begin by noting that the parameters derived from scattered light fitting, such as by Ahmic et al., will be very sensitive to the choice of parametrization of the phase function, which will in turn affect the resulting shape of the vertical profile, its width-radius-line of sight inclination dependences, and the disk radial profile derived. We also note that fits to the sky-projected vertical profile (similar to those carried out on our ALMA data in Fig. 3) by Golimowski et al. 2006 indeed use a combination of two Lorentzians, but Apai et al. 2015 finds an equally good fit using combinations of two Gaussians, as done in our ALMA fits and as physically motivated in Sect. 5.1. This suggests that a double-Gaussian distribution may be equally good at reproducing the scattered light observatio

\subsection{On the SW/NE asymmetry in the mm dust: comparison with the 885 $\mu$m data}
\label{sect:simpledisk_dbgauss_band7}
Independent of the models explored above, the vertically integrated radial profile of the residuals of the 1.33 mm data shows no significant SW/NE asymmetry (bottom left panel of Fig. \ref{fig:combores} and \ref{fig:combores_dbgauss}, where the shaded region is the $\pm1\sigma$ confidence interval). Averaging emission between 30 and 80 au on each side of the star, as in \S\ref{sect:res}, we find a negligible residual flux difference between the NE and the SW of $82\pm85$ $\mu$Jy. Varying the ranges over which we radially average does not affect this result. We conclude that we do not detect a significant SW/NE asymmetry in the new 1.33 mm dataset.

Given the discrepancy with the 15\% asymmetry reported in the 885 $\mu$m dataset \citep{Dent2014}, we independently re-reduced the dataset as described in \S2.2 of \citet{Matra2017a}. Just like the 1.33 mm, compact configuration dataset, this consisted of a mosaic pointed $\pm5\arcsec$ from the stellar location, and along the belt's PA. We imaged the continuum using the \textit{tclean} task (\textsc{CASA} v5.1.0) in multi-frequency synthesis, mosaic mode, using natural weights and multiscale deconvolution to best recover faint, extended emission. We obtained an image with synthesized beam size of $0.70\arcsec\times0.55\arcsec$ at a PA of $64\fdg4$, and measure an RMS noise level of 70$\mu$Jy beam$^{-1}$. The spatially integrated line flux measured on the primary-beam-corrected map as in the 1.33 mm data is $66\pm6$ mJy. 

As done for the 1.33 mm data, we extract the subset of the dataset containing visibilities at $u-v$ distances $>300 k\lambda$ to separate the stellar emission from the disk contribution. We find that the star is not significantly detected. This allows us to set a 3$\sigma$ upper limit of $<786$ $\mu$Jy on the stellar flux at 885 $\mu$m, which is consistent with the expected 196 $\mu$Jy obtained through Rayleigh-Jeans extrapolation from near-IR wavelengths. We therefore include the star with a fixed flux of 196 $\mu$Jy in the modelling below.

We simultaneously fit the visibilities obtained for both pointings as described in \S\ref{sect:modmet}, using the single and double vertical Gaussian models described in \S\ref{sect:simpledisk} and \S\ref{sect:simpledisk_dbgauss}. For each of the two models, we fix the model parameters to the best fit values found for the higher resolution and sensitivity, 1.33 mm dataset (Table \ref{tab:modelcomparison}, Model 1 and 2). We leave as free parameters the scale factor $\rho_0$ governing the total flux of the belt, which is different at the shorter wavelength, as well as an RA and Dec offset and a weight-rescaling factor for each of the two pointings that should be different from the 1.33 mm datasets. 

 \begin{figure*}
 %\vspace{-20mm}
  %\centering
  \hspace{-14mm}
   \includegraphics*[scale=0.39]{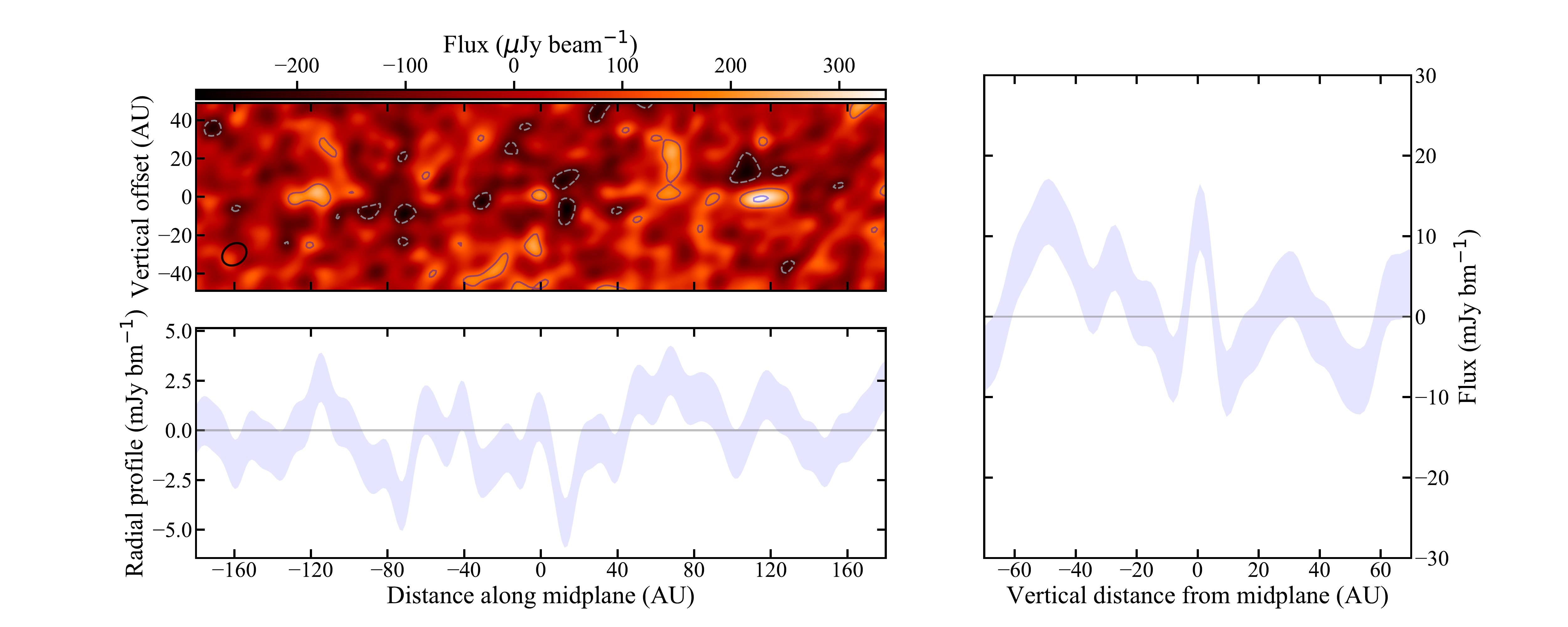}
 \vspace{-6mm}
 \caption{Same as \ref{fig:combores}, but for a fit to the 885 $\mu$m data (see \S\ref{sect:simpledisk_dbgauss_band7}). Despite the lower angular resolution, the residuals show remarkable similarity to those obtained from 1.33 mm data.
 } 
 \label{fig:combores_b7}
 \end{figure*}

 \begin{figure*}
 \vspace{-5mm}
  %\centering
  \hspace{-14mm}
   \includegraphics*[scale=0.39]{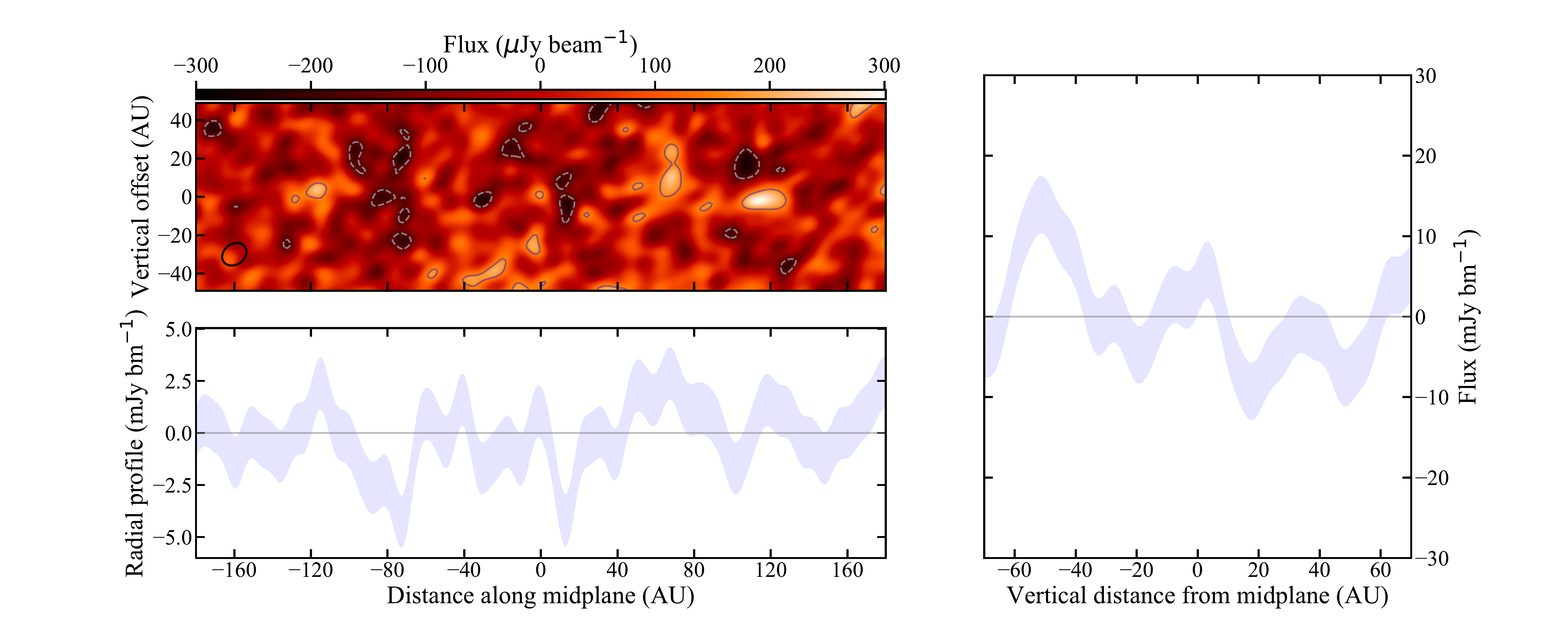}
 \vspace{-6mm}
 \caption{Same as \ref{fig:combores_dbgauss}, but for a fit to the 885 $\mu$m data (see \S\ref{sect:simpledisk_dbgauss_band7}). Despite the lower angular resolution, the residuals show remarkable similarity to those obtained from 1.33 mm data.} 
 \label{fig:combores_dbgauss_b7}
 \end{figure*}

Fig. \ref{fig:combores_b7} and \ref{fig:combores_dbgauss_b7} show residual maps for the single and double vertical Gaussian models, respectively. We find the residuals, particularly the radial and vertical profiles, to be remarkably similar to those obtained for the 1.33 mm observations. In fact, despite the lower resolution and sensitivity, the 885 $\mu$m dataset is also better fitted by a model with a double over a single Gaussian vertical distribution, and also shows no significant SW-NE asymmetry. Averaging emission between 30 and 80 au on each side of the star, as in \citet{Dent2014}, we find a negligible residual flux difference between the NE and the SW sides of $60\pm130$ $\mu$Jy. 

An interesting hypothesis is that the two Gaussian populations have different spectral indices, which could be caused by a difference in the size distribution of grains between the two populations. While this can in principle be tested with resolved observations at different wavelengths, such as presented here, the coarse resolution of the 885 $\mu$m data (which, at 100 au from the star, is of order the scale height of the broad Gaussian population) and the relatively small difference in wavelength between the two datasets do not allow us to test this hypothesis. Future ALMA observations at a S/N and resolution comparable to the 1.33 mm dataset presented here, and at a wavelength as far as possible from 1.33 mm, are needed to explore this idea.
%While no significant large-scale SW/NE asymmetry is seen, it is worth noting that compact excess emission is marginally detected in the residuals of both observations at $\sim$70-80 au from the star, a few au above the midplane, on the belt's SW side (see Fig. \ref{fig:combores_dbgauss} and \ref{fig:combores_dbgauss_b7}, top left panels). However, its sky-projected radial location is larger than the $\sim$60 au centroid of the clump seen in CO \citep{Matra2017a} and \ion{C}{1} \citep{Cataldi2018}. Deeper observations are necessary to confirm and ascertain the nature of this potential excess emission.

\section{Discussion}
\label{sect:disc}

In \S\ref{sect:res} and \S\ref{sect:modelling} we modelled ALMA observations of the $\beta$ Pictoris disk at 1.33 mm and 885 $\mu$m, which show the belt's vertical dust density distribution is much better fitted by the sum of two Gaussian distribution compared to a single Gaussian distribution. % to show that 1) as opposed to the CO, the vertical profile of the mm dust cannot be fit by a single-component Gaussian, with a double-Gaussian distribution producing a much improved fit. 2) The SW/NE asymmetry in the mm dust continuum is weak, if present at all, and not as pronounced as for the CO. 3) The warp observed for $\mu$m-sized grains in scattered light is not as pronounced in the structure of mm grains, tracing the planetesimal population.
In this Section, we will link this vertical profile to the distribution of particle inclinations, and connect it to their dynamical excitation to infer the potential action of large stirring bodies within or exterior to the belt.

\begin{figure*}
%\vspace{-20mm}
 %\centering
 \hspace{-2mm}
  \includegraphics*[scale=0.43]{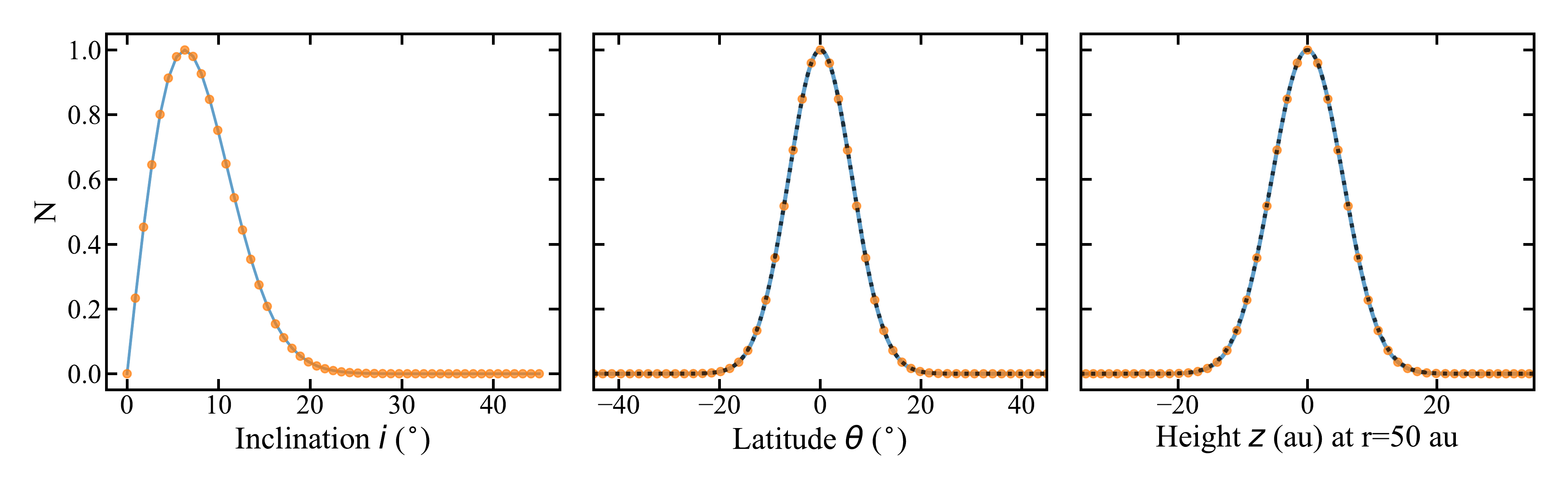}
\vspace{-5mm}
\caption{Conversion from an inclination distribution of particles (left) to a measured latitude (center) and a scale height (right) above the belt midplane. Orange points show the inclination distribution used by \citet{Brown2001} to reproduce the Kuiper belt, how it is well matched (for the small angles considered here) to a Rayleigh distribution (blue line), and how it converts to a density distribution as a function of latitude and height above the midplane that is close to Gaussian (black dotted line).
} 
\label{fig:inctoz}
\end{figure*}

\subsection{Connecting a belt's vertical density distribution to the planetesimals' inclination distribution}
\label{sect:Htoinc}

The observed inclination distributions of planetesimals in the Kuiper belt was first investigated by \citet[][B01 hereafter]{Brown2001}. The authors used a Gaussian multiplied by a sine function to model the observed inclinations of Kuiper Belt Objects (KBOs), and showed that this functional form naturally arises from N-body simulations from gravitational perturbations, or \textit{stirring}, within a planetesimal disk with initially zero inclinations. The function reads $f(i)di\propto\sin{i}\ e^{-\frac{i^2}{2\sigma^2_i}}di$ and for small angles corresponds to a Rayleigh distribution, $f(i)\propto\frac{i}{\sigma_i^2}e^{-\frac{i^2}{2\sigma^2_i}}$. This result is consistent with other dynamical studies also finding that gravitational perturbations of planetesimals in a thin disk lead to Rayleigh distributions of inclinations and eccentricities \citep[e.g.][]{Lissauer1993, IdaMakino1992}. For direct comparison with the Kuiper belt, we here follow the notation of B01 but note that dynamical studies typically express the inclination distribution as a function of the mean of the squares of the inclinations $\langle i^2\rangle$, with $\langle i^2\rangle=2\sigma_{i}^2$ in the expressions quoted above.
Fig. \ref{fig:inctoz} (left) shows an example inclination distribution (following B01) for a population of particles with a mean inclination of $\sim7.9^{\circ}$ (orange points) and shows how this is well matched to a Rayleigh distribution (blue line).

Under the assumption of circular orbits, this inclination distribution of planetesimal orbits $f(i)$ in a belt at a given radius can be linked to the distribution of latitudes $L(\theta)$ (about a planetary system's ecliptic plane) at which the planetesimals are found (B01), through
\begin{equation}
L(\theta)=\int^{\pi/2}_{\theta}f(i)\frac{\cos{\theta}}{\sqrt{\sin^2i-\sin^2\theta}}di.
\label{eq:rawbrown}
\end{equation}
The integral can be simplified through use of the small angle approximation ($\sin{x}\sim x$ for both $i$ and $\theta$) given the small inclination angles considered here. For example, Fig. \ref{fig:inctoz} (centre and right) shows the latitudinal and height 
distribution derived from an inclination distribution (left) with a mean $\bar{i}$ of 7.9$^{\circ}$ and an RMS of 8.9$^{\circ}$, corresponding to the hot population of the $\beta$ Pic belt.
At the height corresponding to 3$\sigma$ of the distribution (16.5 au, implying a latitude of $18.9^{\circ}$), the small angle approximation is accurate to within 2\%.
Taking the small angle approximation, the integral in Eq. \ref{eq:rawbrown} becomes
\begin{equation}
L(\theta)\sim\frac{\cos{\theta}}{\sigma_i^2}\int^{\pi/2}_{\theta}\frac{ie^{-\frac{i^2}{2\sigma^2_i}}}{\sqrt{i^2-\theta^2}}di,
\label{eq:intsimp}
\end{equation}
%with the inclination distribution $f(i)\propto\sin{i}\ e^{-\frac{i^2}{2\sigma^2_i}}$ of \citet{Brown2001} approximating to a Rayleigh distribution,
%$f(i)\sim\frac{i}{\sigma_i^2}e^{-\frac{i^2}{2\sigma^2_i}}$.
and can be solved analytically leading to the expression
\begin{equation}
L(\theta)=\sqrt{\frac{\pi}{2\sigma_i^2}}\cos{\theta}\ \mathrm{erf}\left(\sqrt{\frac{\frac{\pi^2}{4}-\theta^2}{2\sigma_i^2}}\right)e^{-\frac{\theta^2}{2\sigma_i^2}}.
\end{equation}
Given the small inclinations $i$ expected will inevitably lead to small latitudes $\theta$, we can again use the small angle approximation to obtain $\cos{\theta}\sim1$ and $\mathrm{erf}\left(\sqrt{\frac{\frac{\pi^2}{4}-\theta^2}{2\sigma_i^2}}\right)\sim1$ (the latter easily shown given e.g. $\theta\ll5\sigma_{i}$ and small $\theta$).
Then, we come to the conclusion that the latitudinal distribution of particles with a Rayleigh distribution of inclinations (Fig. \ref{fig:inctoz}, left) is simply a Gaussian with standard deviation equal to the $\sigma_i$ parameter of the inclination distribution, i.e. 
$L(\theta)\propto e^{-\frac{\theta^2}{2\sigma_i^2}}$. We verify this in Fig. \ref{fig:inctoz} (centre) by comparing an evaluation of the numerical integral in Eq. \ref{eq:rawbrown} with the Gaussian derived analytically above.

Finally, for small angles the distribution of heights (rather than latitudes) of the planetesimals will also be very well approximated by the Gaussian
\begin{equation}
\rho(z)\propto e^{-\frac{z^2}{2\sigma_z^2}},
\end{equation}
where $\sigma_z=H=hr\sim \sigma_ir$. Thus, using a Gaussian vertical distribution in our model fits of \S\ref{sect:modelling} is justified by the expectation of a Rayleigh distribution of inclinations for a belt whose dynamics is dominated by gravitational stirring from bodies within the belt alone \citep[e.g.][]{Lissauer1993,StewartIda2000}.
%Therefore, belts of planetesimals on circular orbits with a Rayleigh distribution of (small) orbital inclinations will appear to have Gaussian vertical density distributions. 
The aspect ratio of an observed vertical Gaussian distribution can then be used to derive the mean $\bar{i}$ and root mean square $\mathrm{RMS}_{i}=\sqrt{\langle i^2\rangle}$ of the inclination distribution of belt particles, following
\begin{equation}
\bar{i}=\sqrt{\frac{\pi}{2}}h\ \ \ \ \mathrm{and}\ \ \ \ \mathrm{RMS}_{i}=\sqrt{2}h.
\end{equation}

\begin{figure*}
%\vspace{-20mm}
 %\centering
 \hspace{-2mm}
  \includegraphics*[scale=0.43]{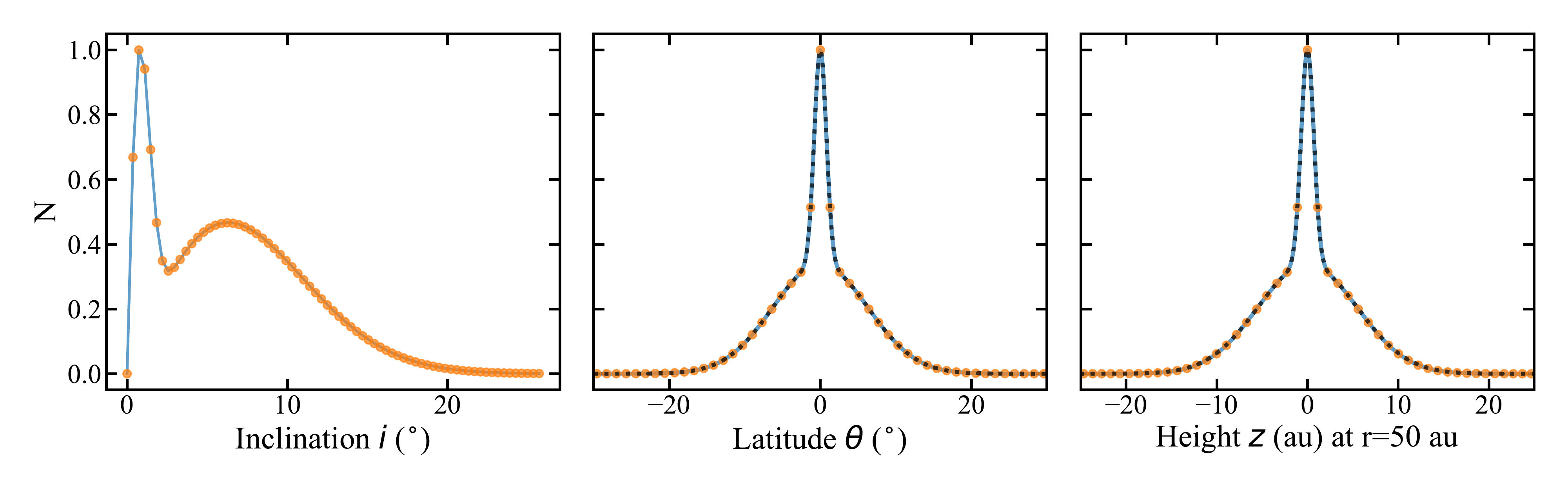}
\vspace{-5mm}
\caption{Same as Fig. \ref{fig:inctoz}, but for a distribution of inclinations that is the sum of two Rayleigh distribution (left). The centre and right panels confirm that this corresponds to a vertical density distribution that is the sum of two Gaussians (black dotted line), as observed for the $\beta$ Pictoris belt in our data.
} 
\label{fig:inctoz_double}
\end{figure*}

In \S\ref{sect:res} and \ref{sect:modelling} we showed how the vertical distribution of mm dust in the $\beta$ Pictoris belt is best described by a sum of two Gaussian distributions. 
%Fig. \ref{fig:inctoz} used as an example the distribution of inclinations $i$ (left) derived for the observed distribution of heights (right) of the broad Gaussian component of best-fit Model 3. In this case, we would use the best-fit model scale height (6.8 au at 50 au radius) to derive a mean inclination of 9.8$^{\circ}$ and an RMS of 11.0$^{\circ}$ for the vertically broad population of planetesimals in the $\beta$ Pic belt.
The derivation above then indicates that the corresponding distribution of particle inclinations is bimodal, corresponding to the sum of two Rayleigh distributions (Fig. \ref{fig:inctoz_double}). From now on, we will refer to these as the \textit{hot} and \textit{cold} populations of particles, with high and low inclinations respectively. The best-fit RMS inclination derived from Model 2 is $0.156^{+0.011}_{-0.009}$ rad ($8.9^{+0.7}_{-0.5}$ degrees) for the hot population and 0.02$^{+0.01}_{-0.01}$ rad ($1.1^{+0.5}_{-0.5}$ degrees) for the cold population.

\subsection{On the origin of the inclination distribution}
\label{sect:origininc}

\subsubsection{External stirring: $\beta$ Pic b}
\label{sect:bpicb}

The only known body which could contribute to structure in the belt is the directly imaged, super-Jupiter mass planet $\beta$ Pic b.
The most recent orbital determination for the planet indicates a best-fit semimajor axis of 8.8 au, an inclination to the line of sight $I_{\rm p}$ of 89.3$^{\circ}$ and a position angle $\Omega_{\rm p}$ of 32.1$^{\circ}$ \citep{Lagrange2018b}. When comparing to the belt as resolved by ALMA, we confirm that the orbital plane of the planet and the belt are misaligned with respect to one another. This is thanks to our high resolution measurement of the belt's position angle $\Omega_{\rm b}$ and inclination to the line of sight $I_{\rm b}$ from thermal emission, which for the first time are free of assumptions on the unknown phase function of the grains, which affected previous determination from scattered light observations. 

The misalignment of the orbital plane of the planet and that of the disk can be characterized by the angle $\mathrm{d}I$, the inclination of the planet orbit with respect to the disk midplane. A simple transformation between reference frames shows that
\begin{equation}
\cos(\mathrm{d}I)=\cos(I_{\rm b})\cos(I_{\rm p})+\sin(I_{\rm b})\sin(I_{\rm p})\cos(\Omega_{\rm p}-\Omega_{\rm b}).
\end{equation}
We find that the belt's inclination $I_{\rm b}$ to the line of sight (see Table \ref{tab:modelcomparison}) for all the best-fit models to be consistent with that of $\beta$ Pic b's orbit \citep[$I_{\rm p}\sim$89.3 deg, but see Fig. 5 in][for an estimate of the uncertainty]{Lagrange2018b}. If confirmed by future observations, that would indicate a planet-belt orbital plane misalignment $\mathrm{d}I\sim2.4$ deg, approximately equal to the difference in position angle between the belt and the planet's orbital plane. %In other words, if the inclinations to the line of sight are confirmed to be the same, the line of nodes of the planet's orbit with respect to the belt's orbital plane is close to perpendicular to the sky plane.

Assuming this misalignment has been retained for the $\sim$23 Myr age of the system, the misaligned orbital plane of $\beta$ Pic b should significantly affect the belt's vertical structure through secular perturbations affecting the belt's planetesimals. In particular, $\beta$ Pic b should impose a forced inclination $i_{\rm p}=\mathrm{d}I\sim2.4$ deg on the belt's particles, and cause their inclinations to oscillate between 0 and 2$i_{\rm p}$ \citep[e.g.][]{Dawson2011, NesvoldKuchner2015}. Particles further out in the belt are affected on longer timescales; this has long been considered the origin of the belt's warp observed in the inner regions out to $\sim$100 au from scattered light observations \citep{Mouillet1997,Augereau2001}, that trace the smallest grains in the collisional cascade.

In this scenario, we would expect 1) the warp to be also present for mm grains, and 2) the vertical distribution of dust density, and hence mm emission, to be enhanced at 0 and 2$i_{\rm p}$ where particles spend most of their time during their oscillations. An example of the expected disk morphology imparted by $\beta$ Pic b is shown in Fig. 1 of \citet{NesvoldKuchner2015}. This shows clear vertical flux enhancements displaced $\sim\pm9$ au from the midplane at 100 au from the star, which are clearly ruled out by the disk spine and vertical profiles derived from the new ALMA data (Fig. 3, central and top panel). Indeed, the mm continuum emission is centrally peaked and displaced by no more than 5 au vertically (at the $3\sigma$ level) throughout the disk midplane. This would limit any sky-projected warp, if present, to less than 2.8 deg (when measured at 100 au), which roughly corresponds to the forced inclination imposed by the planet.

In summary, we find that the disk spine and centrally peaked vertical emission profile are inconsistent with models of the secular perturbations imposed by $\beta$ Pic b on the belt. Taken alone, this could indicate that either 1) the planet is less inclined compared to the belt than currently believed, 2) another massive planet is present in the outer regions of the belt, dominating its dynamics, or 3) $\beta$ Pic b has only recently been put on a misaligned orbit, and has not had time to interact with the outer belt.
Such an alternative picture would have to be reconciled with the scattered light and CO inner disk tilt, and with the hot and cold inclination populations reported here.

\subsubsection{Gravitational stirring within the belt and constraints on the size of the belt's largest bodies}
\label{sect:stirwithin}

The strongest constraint on the double-Gaussian vertical distribution of the belt comes from its radial outer edge, which is least affected by line-of-sight integration of emission originating at different orbital radii. Even if the belt is misaligned with $\beta$ Pic b's orbit, this outer region has not had enough time to be affected by secular perturbations from the planet within the age of the system. Therefore, any structure in this region, and particularly the hot population, is either inherited from the gas-rich protoplanetary phase of evolution \citep[e.g.][]{Walmswell2013}, or must be the result of dynamical excitation from a large body/bodies other than $\beta$ Pic b.

We here consider gravitational stirring by large bodies within the belt itself. Encounters with such bodies will act as a source of dynamical heating, setting the velocity dispersion of particles, and producing a single, Rayleigh distribution of inclinations and eccentricities \citep[see][and references therein]{KokuboIda2012}. Given that the observed inclination distribution of the $\beta$ Pic belt clearly deviates from a single Rayleigh distribution, we can rule out gravitational stirring from within the belt \textit{alone} as the source of the observed vertical structure. Nonetheless, we here consider gravitational stirring as a potential source of either the hot or cold populations, while keeping in mind that other mechanisms are required to produce the other population.
%One possibility is that the inclination distribution results solely from viscous stirring by large bodies within the belt. In this case, gravitational scattering acts as a source of dynamical heating to set the velocity dispersion of particles \citep{IdaMakino1992}. This then sets the width of the inclination distribution and hence the belt's scale height, and can start the collisional cascade itself \citep[e.g.][]{KenyonBromley2001}.
%As mentioned in \S\ref{sect:Htoinc}, such gravitational scattering produces a Rayleigh distribution of particle inclinations; 
%This %corresponds to a triaxial Gaussian distribution (in cylindrical coordinates) for the distribution of random particle velocities, and 

Gravitational stirring leads to a relative velocity of particles of the form $v_{\rm rel}=\sqrt{1.25\langle e^2\rangle+\langle i^2\rangle}v_{\rm Kep}$, where $v_{\rm Kep}$ is the local Keplerian velocity \citep[e.g.][]{Lissauer1993}. Since $\sqrt{\langle e^2\rangle}=2\sqrt{\langle i^2\rangle}$ \citep{IdaMakino1992}, the RMS of the inclination distribution is a direct measure of the relative velocity of particles, following
\begin{equation}
v_{\rm rel}=\sqrt{6\langle i^2\rangle}v_{\rm Kep}=103.9M_{\star}^{0.5}hr^{-0.5},
\end{equation}
where $h$ is unitless, $r$ is in au, $M_{\star}$ is in Solar masses, and the resulting $v_{\rm rel}$ is in km/s.
Then, for an assumed stellar mass of 1.75 M$_{\odot}$ \citep{Crifo1997}, the cold and hot populations in the $\beta$ Pic belt (Model 2) indicate approximate relative velocities of 0.27 and 2.1 km/s at 50 au, dropping to 0.16 and 1.23 km/s at 150 au. %For simplicity, these estimates and the calculations below assume the hot and cold populations can be treated as separate populations, although we note that this may not be the case.

Viscous stirring from a large body (or bodies) will act to increase the velocity dispersion of planetesimals, and consequently their inclinations, over time.
%Given enough time, the velocity dispersion of particles in the belt will increase up to the escape velocity of the largest bodies (REF). Therefore, the relative velocities measured above for the hot population can be used as an estimate of the size of the bodies stirring the hot population; we find that masses of X and Y M$_{\oplus}$ would be needed to excite the hot population at 50 and 150 au in the $\beta$ Pic belt. However, these masses should be considered a lower limit, as they assume that the excitation timescale is shorter than the age of the system. 
Following \citet{IdaMakino1993}, assuming the system has been evolving from a perfectly cold disk ($e,i\sim0$ at $t=0$ Myr) for the age of the system ($t\sim23$ Myr) we can then relate the measured inclinations to the product of the mass and the surface density of the largest bodies. 

\begin{figure}
%\vspace{-20mm}
 %\centering
 \hspace{-14mm}
  \includegraphics*[scale=0.51]{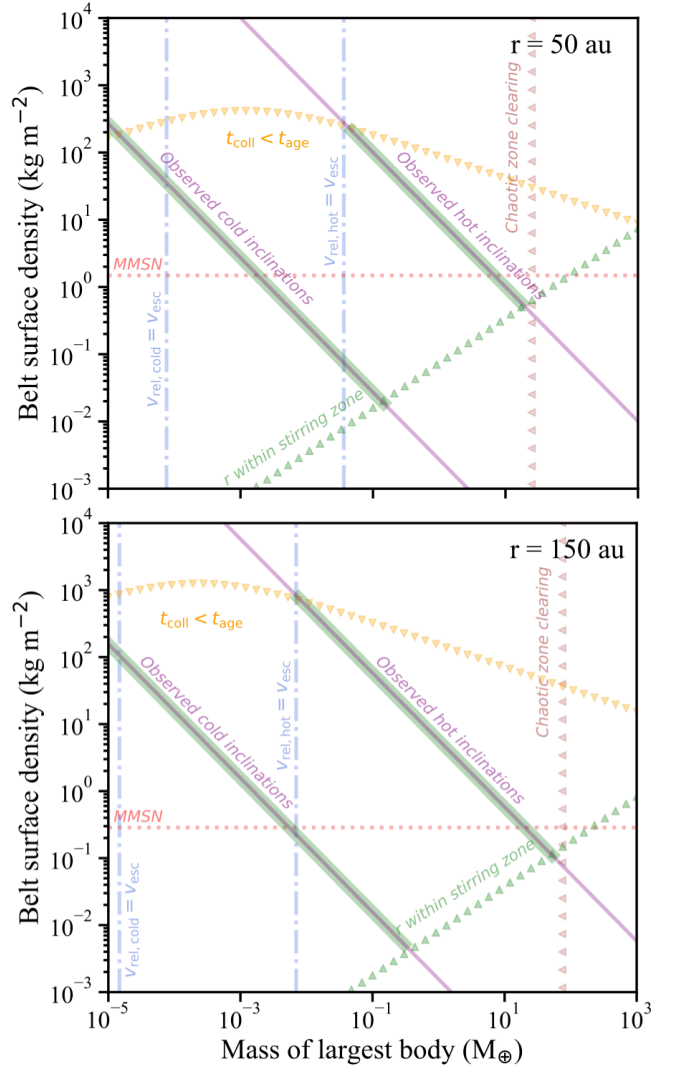}
\vspace{-5mm}
\caption{Dynamical constraints on the mass and surface density of the largest bodies within the belt that could be producing (at 50 or 150 au) the cold or hot population of inclinations observed. The purple lines come from the observed hot and cold population inclination constraints (Eq. 10). The requirement that large bodies have not collided over the age of the system produces the surface density upper limits shown as the orange downward triangles. The intersection between the purple and orange lines represent the lower limit on the mass of the largest bodies. If the stirring bodies are massive enough that $v_{\rm esc}\gg v_{\rm rel}$, this intersection will correspond to the requirement that their escape velocity equals the observed relative velocities of planetesimals in the belt (blue dash-dotted line).
Then, the presence of at least one large body within stirring distance of the radius in question produces the lower limit shown as the green upward triangles. The mass and surface density of the largest bodies are constrained to the portion of the purple line highlighted in green. A further constraint (not affecting the results here) is the lack of radial gaps in the belt's observed radial distribution. This implies that no gap of size equal to the chaotic zone has been opened over the lifetime of the system, setting a mass upper limit (brown leftward triangles in the upper panel). The red dotted line represents the surface density of the Minimum Mass Solar Nebula (MMSN).} 
\label{fig:combodynamics}
\end{figure}

We assume we are in the dispersion-dominated regime and that stirring is dominated by large body/bodies rather than the small planetesimals themselves. Combining Eq. 4.2 and 4.9 of \citet{IdaMakino1993} and solving for the inclination as a function of time, we obtain
\begin{equation}
\sqrt{\langle i^2\rangle}=\left(2C_{\rm I}\frac{M}{M^2_{\star}}r^2\Omega\Sigma t\right)^{\frac{1}{4}}, 
\end{equation}
where the constant $C_{\rm I}\sim2$ \citep{IdaMakino1993}, $\Omega$ is the Keplerian frequency in rad s$^{-1}$, $M_{\star}$ is the stellar mass in kg, $t$ is time in s and $\Sigma$ is the surface mass density of the large stirring bodies (in kg m$^{-2}$), each of mass $M$ (in kg). The measured RMS inclination of a planetesimal population then allows us to set a constraint on the mass times the surface density of the largest bodies of
\begin{equation}
M\Sigma=4.43\times10^3 M_{\star}^{\frac{3}{2}}r^{-\frac{1}{2}}\frac{\sqrt{\langle i^2\rangle}^4}{t}\ \ \mathrm{M}_{\oplus}^2\mathrm{au}^{-2},
\end{equation}
(where $M_{\star}$ is now in M$_{\odot}$, $r$ is in au, and $t$ is in Myr). This leads to a value of $5.9\times10^{-6}$ M$_{\oplus}^2$au$^{-2}$ for the cold population, and 0.022  M$_{\oplus}^2$au$^{-2}$ for the hot population, at 150 au in the $\beta$ Pic belt. This corresponds to the purple lines in the $\Sigma$ vs $M$ space shown in Fig. \ref{fig:combodynamics}.%, where we also consider the case of stirring at 50 au within the belt.

Producing the observed excitation however requires the bodies causing the stirring to not collide (and hence be destroyed) within the age of the system.
The rate of mutual collisions for a population of bodies of equal mass $M$ can be estimated as
\begin{equation}
R_{\rm col}=\sigma n v_{\rm rel}\left(1+\frac{v^2_{\rm esc}}{v^2_{\rm rel}}\right),
\end{equation}
where $\sigma=4\pi R^2$ is the effective cross section of the collision (with $R=(0.24M/\rho)^{1/3}$ being the radius of the large bodies, and $\rho$ their material density), $n$ is the number density of such bodies, and $v_{\rm esc}=\sqrt{GM_{\star}/R}$ their escape velocity. As in Eq. 8, we assume $v_{\rm rel}=\sqrt{6\langle i_p^2\rangle}$ where $\langle i_p^2\rangle$ is the RMS inclination of the large bodies doing the stirring, which is likely primordial and here assumed to be 0.05.
Setting the collision timescale ($t_{\rm c}=R_{\rm col}^{-1}$) to be greater than the age of the system, we can obtain an upper limit to the surface density of the large bodies given their mass, shown as the orange downward triangles in Fig. \ref{fig:combodynamics}. In log-log space, the slope of this surface density upper limit transitions from positive at the low mass end where $v_{\rm esc}\ll v_{\rm rel}$, to negative at the high mass end where $v_{\rm esc}\gg v_{\rm rel}$. 

The intersection between the orange and purple lines indicate the minimum mass necessary for large bodies to produce the observed level of stirring, but at the same time avoid potentially destructive collisions in 23 Myr. It can be shown analytically by combining Eq. 9 and 11 at the high mass end (where $v_{\rm esc}\gg v_{\rm rel}$) that requiring the collision timescale to be equal to the viscous stirring timescale leads to the \textit{planetesimals'} relative velocities to be not exactly equal to, but of order the escape velocity of the largest bodies (blue dash-dotted line). This means that this minimum mass can be approximately estimated by simply setting the planetesimals' relative velocities to be equal to the stirring body's escape velocity.

A further requirement for the surface density of large bodies in the belt is that at least one large body is present at the radii where the stirred population is observed. In other words, if we consider the stirred population near, say, 150 au, there needs to be at least one large body such that its stirring zone encompasses 150 au. This sets a lower limit on the mass surface density (green upward triangles in Fig. \ref{fig:combodynamics}),
\begin{equation}
\Sigma>\frac{M}{2\pi r\Delta r_{\rm stir}},
\end{equation}
where $\Delta r_{\rm stir}=8\sqrt{3}r[M/(3M_{\star})]^{\frac{1}{3}}$ is the width of the stirring zone (equal to twice the large body's feeding zone, following Eq. 3.4 of \citet{IdaMakino1993}).
The intersection of the purple line and green triangles represents the maximum mass for a body to stir a given population and for at least one such body to exist so that its stirred region encompasses a given radius at which the population is observed. 

Finally, we can consider the lack of observed radial gaps in the belt as a further constraint on the large body's mass. This is because a body sufficiently large will open a gap of width roughly equal to its chaotic zone \citep[$\Delta r_{\rm chaos}\sim3r(M/M_{\star})^{\frac{2}{7}}$, e.g.][]{Wisdom1980, MorrisonMalhotra2015, Marino2018b}. Simply setting the width of the chaotic zone to be less than the resolution of our observations leads to an upper limit to the mass of the bodies within the belt. At 150 au, we find this mass to be $<$0.15 M$_{\oplus}$. However, we need to consider the timescale necessary to open this gap. Following Eq. 7 and 8 in \citet{MorrisonMalhotra2015}, the timescale for a 0.15 M$_{\oplus}$ body to open a gap at 150 au in the $\beta$ Pic belt is $\sim5.7$ Gyr. Then, %bodies larger than this may be present and create the observed hot population of inclinations but not open a gap within the age of the system. 
the lack of radial gaps can not only be caused by the chaotic zone not being resolvable, but also by the large body's gap-opening timescale being longer than the age of the system, where the latter sets a more conservative constraint on the maximum planet mass. At 150 au (50 au), we find that a stirring planet should be less massive than 75 (25) M$_{\oplus}$ to not produce observable gaps in 23 Myr (brown left-pointing triangles in Fig. \ref{fig:combodynamics}).

In reality, this estimate is uncertain because the clearing timescale of \citet{MorrisonMalhotra2015} is defined as the `time required to reach 50\% of the final survivor fraction within the cleared zone' and is therefore not to the timescale to produce an \textit{observable} gap. The latter may be shorter, for example, if the observing sensitivity is sufficient to detect gap-belt surface brightness contrasts corresponding to a higher survival fraction of particles in the belt. At the same time, the edge-on viewing geometry would make it harder to detect such gap-belt contrast compared to a face-on belt, potentially making the timescale to produce a detectable gap longer. Detailed N-body simulations combined with radiative transfer and observing simulations are needed to accurately pinpoint this timescale \citep[see e.g.][]{Marino2018b}, but are beyond the scope of this work.
%{\bf I have concerns that the edge-on viewing geometry means that the beam size is not the right constraint on gap detection. The timescale argument is stronger, so maybe the resolution argument is needed?}

Combining the dynamical constraints, we find that the mass of the largest bodies within the belt necessary to stir a hot population such as observed for $\beta$ Pic over 23 Myr should be between 0.007 and 54 M$_{\oplus}$ at 150 au. On the other hand, if we consider viscous stirring as the origin of the cold population, we find that the largest bodies must have masses below 0.4 M$_{\oplus}$ at 150 au (or 0.15 M$_{\oplus}$ at 50 au) to have maintained such a low dynamical excitation. These masses are all far below and therefore consistent with current detection limits from direct imaging \citep{Lagrange2018a}. %The product of the mass by the surface density of these bodies, set by the observed RMS inclination of the hot population (Eq. 10), is 0.022 M$_{\oplus}^2$au$^{-2}$ at 150 au. This indicates surface densities between $\sim$0.34  and $\sim$2500 times the Minimum Mass Solar Nebula (MMSN, scaled by the mass of the $\beta$ Pic star, and shown as the red dotted line in Fig. \ref{fig:combodynamics}). 

To conclude, we underline that viscous stirring by large bodies within the belt alone cannot produce the observed complexity of the vertical structure of the $\beta$ Pictoris belt. However, it could produce either the hot or cold population, while requiring another mechanism to produce the other. If that were to be the case, the observed dynamical excitation of the hot (cold) populations could be produced through viscous stirring by bodies of mass $0.007-54$ ($<0.4$) M$_{\oplus}$ at 150 au.

\subsubsection{Outward migration of a $\beta$ Pic c}
\label{sect:KBcomp}

A bimodal distribution of inclinations such as observed in the $\beta$ Pictoris belt was discovered for our own Kuiper belt (KB) by B01 \citep[after debiasing was taken into account, see also e.g.][]{Kavelaars2009, Petit2011}. 
%This bimodal distribution is retained, although with slightly different parameters, for the Classical KBO population which is found in non-resonant orbits between $\sim40$ and $\sim48$ au. This led to the distinction between the two populations of Hot Classicals (high inclination, HC) and Cold Classicals (low inclination, CC), although it has been shown that the two populations are not clearly distinguished \citep[e.g.][]{Morbidelli2008, VolkMalhotra2011}. Similarly, 
For both belts, we clarify that a single, broad non-Rayleigh distribution of inclinations cannot be excluded \citep[e.g.][]{VolkMalhotra2011}; we just follow the original, bimodal functional form of B01 for direct comparison.
%Scattered KBOs as well as KBOs in resonance with Neptune (e.g. the Plutinos) are also found on high $>5^{\circ}$ inclination orbits, contributing with the HC to what we define here as the \textit{high} inclination population of the KB, which we loosely distinguish from the \textit{low} inclination population which is composed solely of the CC.

\begin{figure}
%\vspace{-20mm}
 %\centering
 \hspace{-2mm}
  \includegraphics*[scale=0.43]{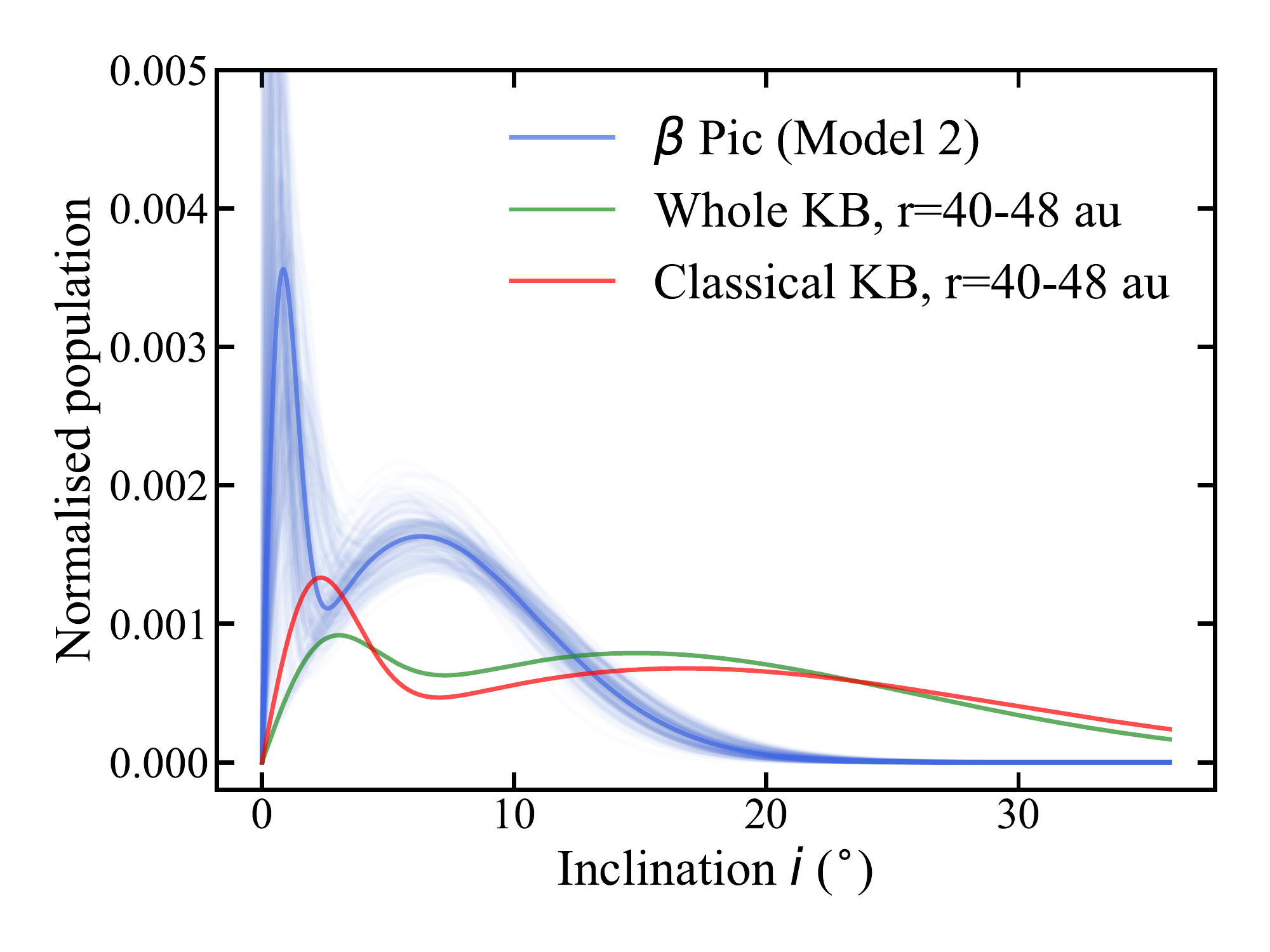}
\vspace{-9mm}
\caption{Comparison between the possible inclination distributions as derived from our observations of $\beta$ Pictoris (blue lines, Model 2) and the Kuiper belt (KB, green and red lines).
} 
\label{fig:incdistcomparison}
\end{figure}

Fig. \ref{fig:incdistcomparison} shows how the inclination distribution of the $\beta$ Pic belt (namely, best-fit Model 2) compares with that of the KB from B01. %Given that the KB is radially narrower than the $\beta$ Pic belt, for $\beta$ Pic we display how the inclination distribution varies at a range of radii. 
Overall, we find lower inclinations for both the hot and cold inclination populations of the $\beta$ Pic belt when compared to the KB, indicating an overall lower level of dynamical excitation.
The fraction of all particles residing in the cold population are however consistent between the $\beta$ Pic belt and the KB ($20^{+5}_{-4}$\% and $\sim19-26$\%), with both belts' mass being dominated by the hot population. %The ratios between the mean inclinations of the cold and hot populations in the $\beta$ Pic system are also similar to our own KB, perhaps indicating that while the level of the excitation may be different, the mechanism to produce this excitation may be similar.

In the Kuiper belt, the high inclinations seen for the hot classicals as well as the resonant and scattered populations set important constraints on its formation scenario \citep[for a review, see e.g.][]{Nesvorny2015}. These populations likely formed in a massive planetesimal disk within 30 au and then got implanted into their present day orbit by outward-migrating Neptune. The high inclinations can be attained through sweeping resonances \citep[e.g.][]{LevisonMorbidelli2003} or scattering by Neptune during its migration \citep{Gomes2003}. \citet{Nesvorny2015} showed that the latter mechanism can work as long as Neptune's migration was slow ($>$10 Myr).

By analogy with the Kuiper belt, we consider whether $\beta$ Pic's structure could be produced by outward migration of a hypothetical $\beta$ Pic c planet located at the inner edge of the belt. The presence of such a planet has been suggested to produce the SW/NE asymmetry observed for the CO gas \citep{Dent2014,Matra2017a} as well as for the small dust grains at wavelengths below 24 $\mu$m \citep[e.g.][]{Telesco2005}, following models predicting resonant planetesimals to concentrate at specific longitudes relative to the planet \citep{Wyatt2003}. 
To produce the asymmetry observed for the CO, two conditions need to be met. First, the 2:1(u) resonance trapping probability (where the (u) indicates planetesimals in 2:1 resonance whose libration center increases from 180$^{\circ}$ during the migration) needs to be sufficiently high, and second, the planetesimal eccentricities need to be excited to a high enough level \citep[we here require e$>$0.3, see Fig. 6, top row of][]{Wyatt2003}.

These eccentricities can be excited by resonant forces, causing them to increase as the radial extent of the migration increases \citep[Eq. 22 in][]{Wyatt2003}. We find that the ratio between the semimajor axis of planetesimals after and before migration should be $\gtrsim1.8$ for eccentricities to reach values above 0.3. For example, outward migration of a planet from 31.4 to at least 57.4 au (close to the inner edge of the belt as observed today) could have swept planetesimals in the 2:1(u) resonance from 50 to at least 91 au and produced sufficiently high eccentricities that an asymmetry may be observed. Given this migration must have taken place for at most the 23 Myr age of the system, we estimate a lower limit on the migration rate of 1.13 au/Myr.

Ensuring the 2:1(u) trapping probability is sufficiently high \citep[we here require it to be above 50\%, see Eq. 8 and Table 1 of][]{Wyatt2003} allows us to link this migration rate to the mass of the migrating planet. We find that a planet with mass of at least $28$ M$_{\oplus}$ must have migrated at least $\sim$25 au to produce an asymmetry as observed for the CO. Therefore, it is possible that outward migration of a super-Neptune sized planet may have produced the SW/NE asymmetry observed for the CO and for the small dust at short wavelengths. Can this however be reconciled with the lack of asymmetry for the mm-sized dust?

\citet{Wyatt2006} showed that the appearance of resonant structure depends on the size of the observed grains, with grains below a critical size $D_{\rm crit}$ falling out of resonance. This size depends solely on the migrating planet's mass \citep[and some stellar parameters, see Eq. 14 in][]{Wyatt2003}. For a planet of at least $28$ M$_{\oplus}$, this size corresponds to 242$\mu$m, although grains up to $\sim$10 times this size can show significantly smeared resonant structure. The observed emission in the ALMA 1.33mm dataset is dominated by grains of size similar to the observing wavelength or a few times smaller \citep[see e.g. Fig. 6, bottom left in][]{Wyatt2003}. It therefore seems plausible that the resonant structure could be smeared out to the extent that it is undetected in the 1.33mm ALMA data as well as in the 24$\mu$m \citep{Telesco2005} and 70-500 $\mu$m \citep{Vandenbussche2010} datasets. Observations at wavelengths even longer than presented here could clarify whether this is the origin of the lack of observed structure at 1.33mm.
%{\bf Could you make a prediction here for the presence of the resonances in observations at longer wavelengths?}

The model predicts the structure of the short lived, smallest grains and the CO to be asymmetric. This means that the asymmetric resonance structure is only lost, or smeared out, at intermediate grain sizes (corresponding to mid to far-IR/mm wavelengths). Asymmetric resonant structure should therefore be seen for the smallest grains, for the CO, and for the large resonant grains of sizes $\gg242 \mu$m.
Note, however, that while the distribution of CO and blow-out grains traces the collisional mass loss rate from the belt (which is proportional to the relative velocity of particles multiplied by the square of their cross section per unit volume), the distribution of the large resonant grains simply traces the particles' cross sections per unit volume.
That has the important implication that the SW-NE asymmetry will be much less pronounced and potentially hardly detectable for large resonant grains when compared to the blow-out grains and the CO. This means that the ALMA 1.33 mm and any longer wavelength data \textit{could} probe resonant grains, but the resonant structure may be too faint for detection.

%The level of asymmetry varies depending on grain sizes \citep{Wyatt2006} as 1) the smallest unbound grains ($\beta>0.5$) are removed from the system soon after preferential collisional production at a given longitude, and therefore should show an asymmetry; 2) the largest grains and planetesimals ($\beta<\beta_{\rm crit}$) are bound and in resonance with the planet, so should also show an asymmetry; while 3) intermediate-sized grains ($\beta_{\rm crit}<\beta<0.5$) fall out of resonance and should not show an asymmmetry.

%The value of $\beta_{\rm crit}$ and hence the grain size $D_{\rm crit}$ at which these transitions occur depend only on the ratio of the mass of the migrating planet to that of the star \citep[Eq. 13 and 14 in][]{Wyatt2006}. In \S\ref{sect:res} we reported no significant asymmetry in 0.88 and 1.3 mm observations of $\beta$ Pictoris, which is consistent with previous results at . This constrains the mass of the migrating planet to $\lesssim1$ M$_{\oplus}$, and its migration rate to $\lesssim$0.08 AU Myr$^{-1}$ (i.e. 1 AU in 13 Myr if we assume the migration started at $\sim$10 Myr). The above migration rate is calculated by requiring the 2:1 resonance trapping probability to be above 50\% \citep[the 2:1 in particular is necessary to produce a SW/NE asymmetry,][]{Wyatt2003}. Given a deprojected centroid of the SW CO clump emission at $\sim$85 AU, assuming this is created from enhanced collisions at the 2:1 resonance, $\beta$ Pic c should be located in the inner regions of the belt, at $\sim$63 au.

As well as the SW/NE asymmetry and the high inclinations observed at the estimated location of the resonant clumps, the outward-migrating planet needs to be able to produce high inclinations all the way out to the outer edge of the disk (i.e. beyond the resonant clumps), as well as the tilt observed for the CO and blow-out grains. A hot population far out in the disk may be produced via outward scattering, as seen for the scattered population of the Kuiper Belt \citep{Brown2001}. At the same time, the inner tilt seen for CO and small grains may be a projection effect caused by two asymmetric, azimuthal resonance clumps in front and behind the plane of the sky in a non perfectly edge-on configuration \citep[see e.g.][Fig. 8, bottom]{Matra2017a}.%In the resonance island (i.e. at the clump location), for the reference model of \citet{Wyatt2003} where a 30 M$_{\oplus}$ planet migrates 45-60 au around a 2.5 M$_{\odot}$ star, relative velocities easily reach the 1-2 km/s required to reproduce the belt's hot population. However, the high inclinations observed by ALMA extend out to (and are most constrained at) the outer edge of the disk (see e.g. Fig. \ref{fig:vertcombo}, top). Then, as well as a resonant population, a \textit{scattered disk} with high inclination particles extending out to 150 au would be necessary to reproduce the data. 

Overall, this $\beta$ Pic c migration scenario could qualitatively reproduce the observed features of the outer belt, but needs to be tested through future planet migration simulations. These should be tailored to reproduce $\beta$ Pic's gas and dust structure as seen by ALMA, and should take into account the influence that $\beta$ Pic b would have on such a planet, which we did not consider here.

\section{Conclusions}
\label{sect:concl}

In this work, we presented new ALMA observations of the $\beta$ Pictoris belt at mm wavelengths. This provided us with the highest resolution picture of large, bound grains in the belt to date, allowing us to study their vertical distribution for the first time.
We report the following findings, confirmed by both image analysis and modelling of the interferometric visibilities from the ALMA data:
\begin{itemize}
\item The vertical distribution of mm grains cannot be fitted by a single Gaussian. In turn, this implies that a single Rayleigh distribution is not a good description of the particles' inclination distribution. We find a model with a bimodal distribution of inclinations, analogous to the Kuiper belt, to be a much better fit to the data. RMS inclinations are $0.156^{+0.011}_{-0.009}$ rad (8.9$^{+0.7}_{-0.5}$ degrees) for the hot component, and $0.02^{+0.01}_{-0.01}$ rad (1.1$^{+0.5}_{-0.5}$ degrees) for the cold component, with the hot component containing $80^{+4}_{-5}$\% of the observed mass.
\item The position angle and inclination of the belt to the line of sight, here determined free of bias from the scattered light phase function, indicate significant misalignment between the belt's and $\beta$ Pic b's orbital planes. Focusing on the belt's inclination to the line of sight, we find that it is consistent with that of $\beta$ Pic b's orbit, which if confirmed would indicate that misalignment is maximum in or close to the plane of the sky. 
\item After fitting an axisymmetric model to the new 1.3 mm and archival 885 $\mu$m visibilities, we find no significant evidence for the previously reported SW/NE brightness asymmetry in the continuum emission arising from mm grains. This is in agreement with previous 24-500$\mu$m dust observations, but in stark contrast with the asymmetries seen in both gas and optical/near-infrared observations.
\item Similarly, we find no significant evidence for a tilt of the inner disk (i.e. a warp) as pronounced as   observed in scattered light and CO observations. While the presence of a smaller tilt is not ruled out, this would now require a mechanism to produce a larger tilt in scattered light and CO gas with respect to the mm grains.
\end{itemize}

We then consider possible origins for the observed inclination distribution observed in the mm grains. We begin by considering the misalignment of $\beta$ Pic b with respect to the belt, which should force the particles' inclination to oscillate between 0 and twice the planet's inclination with respect to the belt plane. This is ruled out by the lack of significant warping and by the vertical distribution observed by ALMA. Furthermore, the strongest constraint on the presence of the hot population comes from the belt's outer edge, which cannot be affected by $\beta$ Pic b's secular perturbations within the age of the system. %We begin by considering radiation pressure, concluding it is unable to produce the observed hot population, and unlikely to act given the large grains probed by ALMA (as testified by the absence of a large radial halo).

We therefore explore alternative scenarios, such as viscous stirring from large bodies within the belt. The latter alone cannot explain the observed inclination distribution, but could be partly responsible for the excitation of either the cold or hot populations.
We lay down a framework to constrain the mass and surface density of large bodies by connecting the observed inclination distribution to the expectation from viscous stirring. We find that producing the hot population would require large bodies of mass between 0.007 and $\sim$54 M$_{\oplus}$ at 150 au within the belt. Most of these bodies cannot produce observable gaps as their gap-opening timescale is longer than the age of the system, although whether the most massive ones $\gtrsim10$ M$_{\oplus}$ could is sensitive to the adopted 'gap opening' definition. On the other hand, keeping the low dynamical excitation of the cold population would require bodies to have masses below 0.4 M$_{\oplus}$ at 150 au.

Another potential explanation could come from the presence of an outward-migrating $\beta$ Pic c at the inner edge of the belt. Such a planet could produce both the SW/NE asymmetry seen in CO and small grains through resonant sweeping, and explain the hot population of inclinations observed in a way analogous to the inferred outward migration of Neptune in the Solar System. A $\gtrsim28$ M$_{\oplus}$ planet having migrated from $\sim$31 to 57 au in the system could produce the SW/NE asymmetry for the CO and small grains. This can likely be reconciled with the lack of a SW/NE asymmetry for grains observed between 24$\mu$m and 1.33 mm, but requires tailored dynamical simulations, for example accounting for the influence of $\beta$ Pic b on such a planet, to draw a more robust conclusion.

%The misalignment of $\beta$ Pic b with respect to the belt should force the particles to the planet's inclination. However, the lack of a significant warp implies that their free - rather than forced - inclination is what determines the width of the observed distribution. Secular perturbations from $\beta$ Pic b itself can stir the inner disk within the age of the system, but not the outer disk and not at a level that is sufficient to reproduce the high inclinations observed.

To conclude, new evidence from ALMA data combined with dynamical arguments indicate that $\beta$ Pic b is unlikely to be the only large body responsible for the belt's observed vertical structure, unless planetesimals were born on high inclination orbits in the earlier protoplanetary disk phase of evolution. Instead, we have shown that the presence of a hot population of inclinations, analogous to the Kuiper belt's, could be produced through dynamical excitation by additional large bodies interior and/or exterior to the belt. Detailed dynamical simulations combined with observations at longer wavelengths or higher resolution are necessary to confirm or rule out the scenarios proposed here.

%\textbf{either planetesimals were born on high inclination orbits, or that a} second planet is present in the system. Given the evidence for asymmetries in the small grain population and in the gas, and drawing the analogy between the observed hot population in $\beta$ Pic's belt and the scattered and resonant populations of the Kuiper belt, we indicate an outward-migrating $\beta$ Pic c at the inner edge of the disk ca as a promising candidate. Detailed dynamical simulations including such planet are needed to confirm this hypothesis, and should be complemented by gas simulations to ascertain the origin of the \ion{C}{1} gas asymmetry.

%we conclude that the inclination distribution of the $\beta$ Pictoris belt cannot be fitted by a single Rayleigh distribution, as expected from an unperturbed disk of planetesimals. Instead, we find an inclination distribution that is similar to the Kuiper belt, and potentially bimodal; the relative fractions of the 


%% Included in this acknowledgments section are examples of the
%% AASTeX hypertext markup commands. Use \url without the optional [HREF]
%% argument when you want to print the url directly in the text. Otherwise,
%% use either \url or \anchor, with the HREF as the first argument and the
%% text to be printed in the second.

\acknowledgments
LM acknowledges support from the Smithsonian Institution as a Submillimeter Array (SMA) Fellow. GMK is supported by the Royal Society as a Royal Society University Research Fellow. This paper makes use of ALMA data ADS/JAO.ALMA\#2012.1.00142.S. ALMA is a partnership of ESO (representing its member states), NSF (USA) and NINS (Japan), together with NRC (Canada), NSC and ASIAA (Taiwan), and KASI (Republic of Korea), in cooperation with the Republic of Chile. The Joint ALMA Observatory is operated by ESO, AUI/NRAO and NAOJ. 
%We are grateful to V. Barger, T. Han, and R. J. N. Phillips for
%doing the math in section~\ref{bozomath}.
%More information on the AASTeX macros package is available \\ at
%\url{http://www.aas.org/publications/aastex}.
%For technical support, please write to
%\email{aastex-help@aas.org}.

%% facilities. A common set of keywords will make these types of searches>

%% copy editing.  Individual instruments can be provided in parentheses,
%% after the keyword, but they will not be verified.

\facility{ALMA}.

\bibliographystyle{apj}
\bibliography{lib}

%% The following command ends your manuscript. LaTeX will ignore any text

\end{document}